\documentclass[aps,preprint,superscriptaddress,nofootinbib]{revtex4-1}
\usepackage{amsfonts}
\usepackage{amsmath}
\usepackage{amssymb}
\usepackage{graphicx}
\usepackage{xcolor}
\usepackage{epstopdf}
\usepackage{epsfig}
\newcommand{\be}{\begin{equation}}
\newcommand{\ee}{\end{equation}}
\newcommand{\bea}{\begin{eqnarray}}
\newcommand{\eea}{\end{eqnarray}}
\newcommand{\ba}{\begin{array}}
\newcommand{\ea}{\end{array}}

\def \nn {\nonumber}
\newcommand{\eq}[1]{(\ref{#1})}

\begin{document}

%%%%%%%%%%%%%%%%%%%%%%%%%%

\preprint{}
\title{Refined Holographic Entanglement Entropy for the AdS Solitons and AdS black Holes}

\vfill

\author{Masafumi Ishihara\footnote{masafumi.ishihara@gmail.com}}
\affiliation{Department of Electrophysics, National Chiao-Tung University, Hsinchu, Taiwan}

\author{Feng-Li Lin\footnote{linfengli@phy.ntnu.edu.tw}\footnote{On leave from National Taiwan Normal University.}}
\affiliation{Department of Physics, Massachusetts Institute of Technology, Cambridge, Massachusetts 02139, USA}
\affiliation{Department of Physics, National Taiwan Normal University, Taipei, 116, Taiwan}

\author{Bo Ning\footnote{ningbo@ntnu.edu.tw}}
\affiliation{Department of Physics, National Taiwan Normal University, Taipei, 116, Taiwan}

%%%%%%%%%%%%%%%%%%%%%
%  Abstract
%%%%%%%%%%%%%%%%%%%%%
\vfill
\begin{abstract}
We consider the refinement of the holographic entanglement entropy for the holographic dual theories to the AdS solitons and AdS black holes, including the corrected ones by the Gauss-Bonnet term. The refinement is obtained by extracting  the UV-independent piece of the holographic entanglement entropy, the so-called renormalized entanglement entropy which is independent of the choices of UV cutoff.  Our main results are (i) the renormalized entanglement entropies of the AdS$_{d+1}$ soliton for $d=4,5$ are neither monotonically decreasing along the RG flow nor positive definite, especially around the deconfinement/confinement phase transition; (ii) there is no topological entanglement entropy for AdS$_5$ soliton even with Gauss-Bonnet correction; (iii)  for the AdS black holes, the renormalized entanglement entropy obeys an expected volume law at IR regime, and the transition between UV and IR regimes is a smooth crossover even with Gauss-Bonnet correction; (iv) based on AdS/MERA conjecture, we postulate that the IR fixed-point state for the non-extremal AdS soliton is a trivial product state.
\end{abstract}
%%%%%%%%%%%%%%%%%%%%%

\maketitle

\tableofcontents

\*\\
\section{Introduction}

   Quantum entanglement is an important theoretical probe to understand some particular feature of the strongly coupled systems \cite{Amico:2007ag,Calabrese:2009qy}, such as the topological ordered phases which are believed to be related to the long-range entanglement \cite{Levin:2006zz,Kitaev:2005dm}. On the other hand, the nature of short-range entanglement for generic ground states yields the famous area law \cite{Srednicki:1993im,Eisert:2008ur}. The entanglement entropy is plagued by the UV cutoff, however, some of the encoded information is related to the counting of number of degrees of freedom. The famous example is the entanglement entropy of the $(1+1)$-dimensional conformal field theory (CFT), for which the coefficient of the logarithmic UV divergent term is proportional to the central charge of the CFT. Thus, one task for the physical interpretation of the entanglement entropy is to extract such kind of the UV-independent piece, or so-called the renormalized entanglement entropy. If these renormalized entanglement entropies are indeed related to the number of the effective degrees of freedom \cite{Myers:2010xs,Myers:2010tj,Myers:2012ed}, then one may expect that they should obey some sort of C- or F-theorem along the renormalization group (RG) flow, at least for relativistic quantum field theories \cite{Zamolodchikov:1986gt,Komargodski:2011vj,Jafferis:2011zi}.

   However, it is difficult to evaluate the entanglement entropy directly even in the text of free field theory, which is usually based on replica method \cite{Callan:1994py,Holzhey:1994we}, not mentioning to evaluate it directly for the strongly coupled theory. Fortunately, it was proposed in \cite{Ryu:2006bv,Ryu:2006ef,Nishioka:2009un} that in the context of AdS/CFT correspondence, the holographic entanglement entropy has a simple geometric representation, which is the area of the minimal hyper surface in the bulk with its UV boundary coincident with the entangling surface in the dual field theory. As usual, the holographic entanglement entropy is plagued by the UV cutoff, and one should be careful to extract the UV-independent piece which is free of the UV cutoff ambiguity. The explicit calculation of the entanglement entropy of the $d$-dimensional conformal field theory (CFT) with its holographic dual the gravity in $(d+1)$-dimensional anti-de Sitter (AdS$_{d+1}$) space, gives the following generic UV scaling structures \cite{Nishioka:2009un}
\be\label{SUV}
S^{(d)}_{\rm UV}\sim
\begin{cases} \frac{R^{d-2}}{\epsilon^{d-2}} + \cdots + \frac{R}{\epsilon} + {\rm const} + {\epsilon \over R}\cdots, & \qquad {\rm d \; odd}, \cr
\frac{R^{d-2}}{\epsilon^{d-2}} + \cdots+ \frac{R^2}{\epsilon^2} + {\rm const}\; \log{\frac{R}{\epsilon}} + {\epsilon^2 \over R^2} \cdots, & \qquad {\rm d \; even},
\end{cases}
\ee
where $R$ is the linear size of the entangling surface, and $\epsilon$ is the UV cutoff. This UV structure is consistent with the one obtained from the fact that the entanglement entropy should be an even function of extrinsic curvature of the entangling surface \cite{Grover}. Moreover, the constant parts in the above are UV-independent pieces, which will not change under the redefinition of the UV cutoff $\epsilon$, and should be identified as the renormalized entanglement entropies.

   Recently, it is proposed in \cite{Hertzberg:2010uv,Liu:2012ee} how to extract from \eq{SUV} the renormalized entanglement entropy. The basic idea is to construct some $d$-dependent function $f_d(R\partial_R)$ of differential operator $R \partial_R$ so that when acting on \eq{SUV} by this operator one will extract the aforementioned UV-independent pieces, which are related to the central charges of the CFTs and should be positive. The detailed form of $f_d$ is given in \cite{Liu:2012ee}. One then applies the same differential operator $f_d$ to the entanglement entropy of the non-CFTs and extracts the corresponding UV independent pieces, which should be the C-functions and are expected to be monotonically decreasing  along the RG flow as $R$ increases. Similar works have recently been done in \cite{Casini:2012ei,Klebanov:2012yf}.
  
     However, the way of extracting the renormalized entanglement entropy is far from unique as in the usual case for other renormalized quantities plagued by UV divergence. Despite that, for the extracting quantities to be related to the number of the underlying degrees of freedom, we should require it to be positive and obey some C-theorem at least at the very beginning of RG flow. The aforementioned $f_d(R\partial_R)$ is devised to satisfy these constraints, and is succinct and scale-adaptive.

       In this paper, we would like to generalize the above extraction scheme to the one for the holographic dual non-CFTs which are gapped or finite temperature version of CFTs, and explore the RG flow behavior of the resultant renormalized entanglement entropies. Our motivation for considering such cases is partly to see if the extraction scheme is universal or not. On the other hand, there may have topological order for gapped systems, which can be encoded in the constant piece of the renormalized entanglement entropy, the so-called topological entanglement entropy \cite{Levin:2006zz,Kitaev:2005dm,Grover}. We will like to examine its existence by the aforementioned extraction scheme for the holographic duals considered here \footnote{ For earlier studies on the UV structure of the holographic entanglement entropy for the AdS solitons, see \cite{Nishioka:2006gr,Klebanov:2007ws,Pakman:2008ui,Ogawa:2011fw}}.

       The holographic duals of the finite temperature version of CFTs are the black holes in AdS spaces, the UV divergence structure of which is similar to \eq{SUV}, thus we can apply the same differential operator  given in \cite{Liu:2012ee} denoted by $f^{(LM)}_d(R\partial_R)$ to extract the renormalized entanglement entropy. As we shall see, the renormalized entanglement entropy shows an expected smooth crossover from the UV regime to the volume law in the IR regime \cite{Swingle:2011mk}, the latter captures the extensiveness of the thermal entropy encoded by the black hole horizon as the entangling surface.

       On the other hand, the holographic duals of the gapped systems considered here are the so-called AdS solitons, which can be obtained by double Wick rotation of the AdS black hole metric and then by compactifying one of the transverse dimensions. The warped size of the compact circle shrinks to zero at some finite value of AdS radial coordinate so that it caps out the rest of the original AdS geometry. The capped geometry implies an IR fixed point of the dual deformed CFT at finite energy scale by the UV-IR correspondence, thus it is dual to a gapped system. Moreover, this compact dimension is a spectator for the dual deformed CFT, i.e., the entangling surface wraps over it, it then yields different UV scaling structure from \eq{SUV}. Instead, for AdS$_{d+1}$ soliton it looks like
\be\label{SUV-1}
{L_{\theta} \over R} S_{\rm UV}^{(d)}
\ee
where $L_{\theta}$ is the fixed proper size of the compact circle, and $S_{\rm UV}^{(d)}$  is the UV structure of AdS$_{d+1}$ given in \eq{SUV}.   We shall then adopt a differential operator to extract the UV-independent piece of the entanglement entropy, denoted as $S_{\rm UV-ind}$, also called the ``renormalized entanglement entropy" for short. Since the two UV scaling structures are related, it is straightforward to see that the differential operator
\be\label{gsubs}
g_d(R\partial_R):= {1\over R} f_d^{(LM)} (R \partial_R) R
\ee
will retain the salient feature of $f^{(LM)}_d$ proposed in \cite{Liu:2012ee}, namely, being succinct and scale-adaptive, and to result in a positive C-function obeying C-theorem at the UV regime.  For example, in the UV limit, the renormalized entanglement entropy extracted from AdS$_{d+1}$ soliton is $c_d/R$ where $c_d$ is the renormalized entanglement entropy extracted from pure AdS$_{d+1}$ space (central charge of the dual CFT).  After some manipulations, we can write down the RG flow of the renormalized entanglement entropy as follows
\be\label{diffsubs}
{dS_{\rm UV-ind} \over dR}=\begin{cases} \frac{1}{(d-2)!!} \,(R \frac{d}{d R} + 1) (R \frac{d}{d R} - 1) \cdots
(R \frac{d}{d R} - (d-4)) \,\frac{d S}{d R}, &
\qquad {\rm d \; odd}, \cr
\frac{1}{(d-2)!!} \,(R \frac{d}{d R} + 2) R \frac{d}{d R} \cdots (R \frac{d}{d R} - (d-4))\, \frac{d S}{d R}, &
\qquad {\rm d \; even},
\end{cases}
\ee
where $S$ is the holographic entanglement entropy for the AdS$_{d+1}$ soliton, whose UV scaling behavior is related to the one for AdS$_{d+1}$ space via \eq{SUV-1}.
Despite the nice feature in the UV-regime, we may not expect the C-theorem to be held along the RG flow due to the nature of the gapped systems, since in higher dimensions the universal part of the entanglement entropy is sensitive to the shape of the entangling surface, and for the AdS soliton case the entangling surface acquires different topology from the pure AdS case due to the compact dimension. This makes the evaluation of the RG flow of the renormalized entanglement entropy for the gapped systems an interesting task. It is indeed the main goal of this paper.

%%%%%%%%%%%%%%%%%%%%%%%%%%%%%%%%%%%%%%%%
\begin{figure}[htbp]%[H]
%\vspace{.3cm}
\includegraphics[width=15cm,height=13cm]{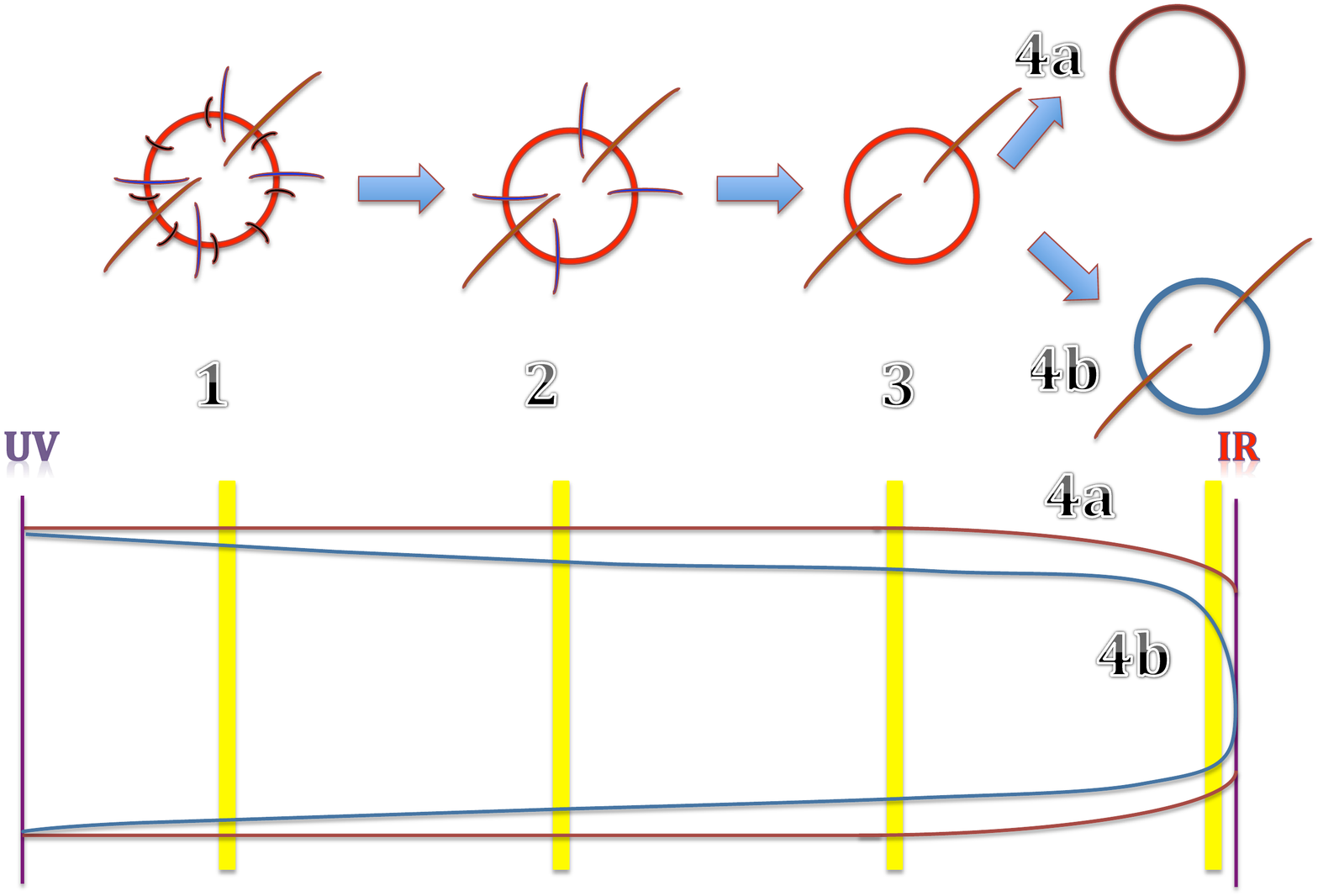}
\caption{Upper: The procedure of MERA or equivalently quantum state renormalization group transformation for the gapped system. The circle at each step denotes the surface enclosing the chosen region, and the  links crossing it denote the entangled pairs which contribute to the entanglement entropy after tracing out the wave function outside/inside the circle.  The length of the link is the distance between the entangled pair, and signifies the entanglement at that length scale.  At each step of MERA, the entanglements at the corresponding scale are removed.  There are two possible end states at the IR fixed-point: (4a) the trivial product state  and (4b) the entangled state  protected by symmetry or topological order.  Lower: The corresponding holographic minimal surfaces in the bulk AdS soliton. The (4a) and (4b) in MERA yield the minimal surfaces of cylinder and disk topologies, respectively. Moreover, the entanglement entropy at each scale of MERA is encoded in the area of the minimal surface above the yellow bar at that scale. As seen, such area for (4a) is negligible compared to (4b). It then suggests that (4a) is a product state without entanglement but (4b) is not. More detailed explanation will be given in section 5.} \label{Fig1}
\end{figure}
%%%%%%%%%%%%%%%%%%%%%%%%%%%%%%%%%%%%%%%%%%

     Besides, if there is no topological order, one may expect the IR fixed point of the gapped system will be a trivial product state after performing proper local unitary transformation to remove the short range entanglement. If so, it implies that the rate of change of the renormalized entanglement entropy along the RG flow approaches to zero in the IR limit. One may also reveal this kind of feature geometrically in the holographic dual gravity.  We will try to argue this is indeed the case based on the proposal of AdS/MERA (multi-scale entanglement renormalization ansatz)  \cite{Swingle:2009bg,EvenblyVidal} by just looking into the dominant topology of the large holographic entangling hypersurfaces.  We briefly  summarize the idea of AdS/MERA and the associated entangled nature of  IR fixed-point state in Fig. \ref{Fig1}, and the more detailed explanation will be given in section \ref{MERAsec}.

     Our paper is organized as follows. In section \ref{solitonsec}, we will extract the UV-independent piece of the holographic entanglement entropy for the AdS$_{d+1}$ soliton with generic form of metrics. Then, we will evaluate numerically the RG behavior of the UV-independent piece. We also discuss how to extract the topological entanglement entropy from the UV-independent piece. In section \ref{AdSBHsec}, a similar consideration goes for AdS black holes.  In section \ref{GBsec}, we will extract the UV-independent piece of the entanglement entropy and its RG flow for the AdS$_5$ soliton and black hole corrected by the Gauss-Bonnet term.  We then conclude our paper in section \ref{MERAsec} by discussing the entangling nature of the IR fixed-point state of the holographic dual theory based on the proposal of AdS/MERA.

\section{Holographic renormalized entanglement entropy for AdS solitons}\label{solitonsec}

   In this section, we will first discuss how to extract the UV-independent piece of the entanglement entropy for the AdS soliton, which is free of the UV cutoff and the associated ambiguity. Then we will discuss how to extract the topological entanglement entropy from the UV-independent piece, which should be encoded in the constant piece in its IR limit.

  We will consider the AdS soliton with following form of metrics in the Poincare coordinates, which can be obtained from the double Wick rotation of some asymptotically AdS space:
\be\label{metric1} ds^2={L^2_{AdS}\over z^2 }\left({dz^2\over
f(z)}+f(z) d\theta^2-dt^2+ dr^2 +r^2 d\Omega_{d-3}\right), \ee where
the harmonic function $f(z)$ can take the general form as follows
\be f(z)=\left(1-k_1 {z\over z_0}\right)\left(1-k_2 {z\over
z_0}\right) (1+\sum_{n=1} c_n z^n ). \ee We assume the $c_n$'s are
chosen appropriately such that $1+\sum_{n=1} c_n z^n$ does not
contain poles and zeros at $z=z_0$.  The parameters $k_1$ and $k_2$
can be tuned to yield different IR behaviors. The metrics include the
pure AdS space by choosing $k_1=k_2=c_n=0$.

 The simplest AdS soliton is the one with $k_1=1$ and $k_2=-1$ and with $c_n$ chosen so that $f(z)=1-({z\over z_0})^{8-d}$.  By choosing the proper period of $\theta$-coordinate, denoted by $L_{\theta}$ to remove the conical singularity, this metric has a smooth tip at $z=z_0$ which corresponds to the IR gap of the dual theory.  Note that the proper size $\sqrt{g_{\theta \theta}} L_{\theta}$ of the $\theta$-direction depends on the RG scale $z$ so that it yields a $d$-dimensional UV theory but a $(d-1)$-dimensional IR theory since the proper size of $\theta$ shrinks to zero there.   One can also turn on some deformation operators to the dual boundary theory of pure AdS soliton, which are encoded in $c_n$'s capturing the deviation from the ones for $f(z)=1-({z\over z_0})^{8-d}$. For example, one can double Wick rotate the AdS$_5$ charged black hole with the harmonic function $f(z)=1-m z^4+q^2 z^6$.  This is then dual to a boundary theory with non-zero current density condensate or magnetic fluxes. More complicated case can be obtained from other deformations of the pure AdS metric, such as the hairy scalar AdS black hole \cite{Hartnoll:2008vx,Horowitz:2010jq} or even AdS R-charged black hole \cite{Cvetic:1999xp}.

   For simplicity, we will set $L_{\rm AdS}=1$ and focus on $d=4$ and $d=5$ case, but also including $d=3$ case for completeness.  Here we refer $d$ to the space-time dimension of the UV theory. In some literature, it refers instead to the space-time dimension of the IR theory, which is one dimension less than the UV one.

\subsection{Extracting the renormalized entanglement entropy}

We choose the entangling surface to be specified as follows by the coordinates $z=0,\, r=R$ with the spatial coordinates of the world-volume: $0\le \theta \le L_\theta\,$ and  $\Omega_{d-3}$. It then has the geometry $S^1 \times S^{d-3}$.
   To evaluate the holographic entanglement entropy, one should find out the minimal surface with its boundary enclosing the entangling surface. This is done by finding the solution of the equation of motion derived from the action for the area of the above hyper-surface, i.e.,
\be\label{Action} A=\int \sqrt{\det g_{\rm ind}}=
\Omega_{d-3} \int_{\epsilon}^{z_m} dz \; {r^{d-3} \over z^{d-1}} \sqrt{1+f
\dot{r}^2}:= \Omega_{d-3} \int_{\epsilon}^{z_m} dz \; \mathcal{L},
\ee
where $g_{\rm ind}$ is the induced metric on the hyper-surface, and
$\dot{r}={dr \over dz}$. The holographic entanglement entropy $S$ is related to the area $A$ by $S = \frac{L_{\theta}}{4 G_N} A$.
\footnote{ For simplicity, hereafter we will omit the angular factor $\Omega_{d-3}$
and will not distinguish between $A$ and $A/\Omega_{d-3}$ and similarly for the quantities related to $A$ such as $S$,
$S_{\rm finite}$ and $S_{\rm UV-ind}$. }

\begin{figure}[h]
\begin{center}
\includegraphics[scale=1.10]{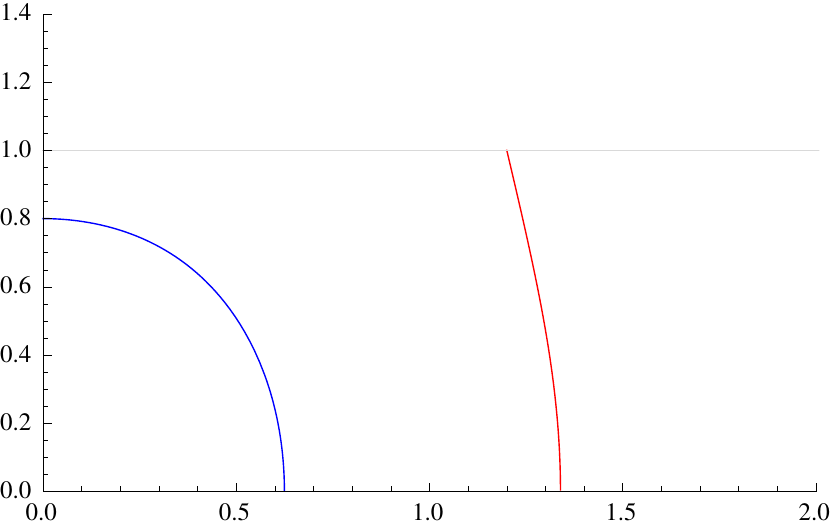}
\end{center}
\caption{Disk (blue) and cylinder (red) topology of the minimal surface for AdS soliton.}
\label{figtopologies}
\end{figure}

The equation of motion for $r(z)$
explicitly is \be\label{EOM} 2(d-1) f^2 r \dot{r}^3+2z(d-3- r
\dot{r} \dot{f})+f(2(d-3) z\dot{r}^2-r (-2(d-1) \dot{r}+z
\dot{f}\dot{r}^3+2z \ddot{r}))=0, \ee where $\dot{f}={df(z)\over
dz}$.  The minimal surface will have different IR behaviors
depending on the linear size $R$. For generic AdS soliton metric,
the small $R$ minimal surface will have a disk topology and $z_m$ is
the turning point such that $r(z_m)=0$. On the other hand, the large $R$ one will
end on the $z=z_0$, thus $z_m=z_0$ with a cylinder topology, see
Fig.\,\ref{figtopologies}.  However, for the case with extremal
harmonic function, i.e., $k_1=k_2=1$, only exists disk topology for
all $R$.

   Varying $A$ with respect to $R$ with $z=\epsilon$ fixed, and using  the Hamilton-Jacobi method, we find that \cite{Liu:2012ee}
\be\label{aRG}
{d A \over d R}= -\mathcal{H}(z_m) {dz_m\over dR}-\Pi(\epsilon){d r(\epsilon)\over dR}=-\Pi(\epsilon){d r(\epsilon)\over dR},
\ee
where
\be
\Pi:={\delta \mathcal{L} \over \delta \dot{r}}={r^{d-3} f \dot{r} \over z^{d-1} \sqrt{1+f \dot{r}^2} }, \qquad \mathcal{H}=\Pi \dot{r} -\mathcal{L}=-{r^{d-3}  \over z^{d-1} \sqrt{1+f \dot{r}^2} }.
\ee
The first term in the first equality of \eq{aRG} is dropped because of the IR boundary condition for the minimal surface, i.e.,
\bea
 r(z_m)=0 \;\; \mbox{s.t.}\;\; \mathcal{H}(z_m)=0 && \qquad \mbox{for disk topology},
\\
 {dz_m\over dR}={dz_0 \over dR}=0 &&  \qquad \mbox{for cylinder topology}.
\eea Note that ${dA\over dR}$ only depends on the UV behavior of the
solution $r(z)$. So the resulting scaling behavior should hold for
both disk and cylinder topologies. However, since the UV boundary
condition alone cannot determine the full solution, some IR
information will be encoded in $r(z)$ and affect the renormalized entanglement entropy implicitly.

   Therefore, we only need to extract the UV behavior of the solution $r(z)$ to yield ${dA\over dR}$, from which we can obtain the RG flow of the holographic entanglement entropy after subtracting off the UV divergence and its associated ambiguity.  We postulate the UV behavior of the solution $r(z)$ as
\be r(z)=R+ b_0 \log {z\over R} + \sum_{n=1} (a_n+b_n \log {z \over R}) z^n\,,
\ee
we then plug it into \eq{EOM} to determine $a_n$'s and $b_n$'s.

\subsubsection{AdS$_5$ soliton}

For concreteness, we consider $d=4$ case first. We find that
\be\label{rz4d} r(z)=R-{z^2\over 4R}+a_4(R) z^4 +{ (c_1-k_1-k_2)z^3
\over 6 R z_0} + ({z^4 \over 32 R^3}-{(c_1-k_1-k_2) z^5 \over 40 R^3
z_0}) \log{z \over R}+\cdots, \ee where $\cdots$ denotes the higher
order terms which can be determined by $a_4$, $k_i$'s and $c_n$'s
but are not relevant for our purpose. An important point is that the
equation of motion at the UV expansion can not determine $a_4(R)$.
Instead one should determine it by solving the full equation of
motion. In other word, $a_4(R)$ encodes some IR information of the
minimal surface and the nontrivial RG flow of the holographic
entanglement entropy. Especially, it should tell when  the phase
transition occurs between disk and cylinder topologies by tuning $R$. This phase transition is nothing but the deconfinement/confinement phase transition \cite{Witten:1998zw} with disk topology corresponding to deconfined phase at small $R$, and the cylinder one to the confined phase at large $R$ \cite{Nishioka:2006gr,Klebanov:2007ws}.

   Plugging \eq{rz4d} into \eq{aRG}, we obtain
\be\label{dadrads5} {d A \over dR}=-4 R a_4(R) +{-k_1^2-k_2^2-k_1 k_2 +c_1 (k_1+k_2)
-c_1^2+c_2  \over 2 z_0^2}-{3\over 32 R^2}+\;\mbox{UV-divergent terms}\;
+\mathcal{O}(\epsilon), \ee
where $\mathcal{O}(\epsilon)$ terms vanish at $\epsilon \to 0$ limit and are not relevant. The
UV-divergent terms are \be\label{uvdivads5} {1\over 2 \epsilon^2}-{1\over
8 R^2} \log({\epsilon \over R})  \ee
which are only defined up to the redefinition of the UV cutoff $\epsilon$.  For example, redefining $\epsilon$ by $a_0 \epsilon(1+a_1 \epsilon + \cdots)$ will then shift \eq{uvdivads5} by terms of $\mathcal{O}(R^0)$ and $\mathcal{O}(R^{-2})$ with finite UV cutoff-dependent coefficients. This means that the terms of $\mathcal{O}(R^0)$ and $\mathcal{O}(R^{-2})$ in \eq{dadrads5} are not universal but depend on the UV cutoff. To obtain a UV cutoff-independent refinement of the holographic entanglement entropy, i,e, the renormalized entanglement entropy, we shall then subtract these kind of terms from \eq{dadrads5}.

   In this paper, we will consider the differential subtraction scheme given in \eq{gsubs} and \eq{diffsubs}. To demonstrate how \eq{diffsubs} is arrived, we take the current example, i.e., $d=4$. Using $f_4(R\partial_R)$ given  in \cite{Liu:2012ee} and \eq{gsubs}, after some manipulations the differential operator acting on $A$ to extract $S^{(4)}_{\rm UV-ind}$ is
\be\label{oprtSAdS5}
g_4(R\partial_R)={1\over 2} (R\partial_R +1)(R\partial_R-1).
\ee
From $S^{(4)}_{\rm UV-ind}:=g_4(R\partial_R) A$ and using the commutator relation $[\partial_R, R \partial_R]=\partial_R$, we can then obtain
\be\label{RGAdS5}
{d S^{(4)}_{\rm UV-ind} \over d R}={1\over 2} (R\partial_R +2) ( R \partial_R ) {d A\over dR}.
\ee
This is the $d=4$ case in \eq{diffsubs}.

Usually in higher dimensions, the renormalized entanglement entropy is determined not only by the intrinsic geometry of the entangling surface, but also its embedding in the spacetime, and in particular related to trace anomaly \cite{Solodukhin:2008dh}. For $d=4$ QFT there are two kinds of anomalies, related to the Euler density (A-type) and the square of the Weyl tensor (B-type) of the entangling surface, respectively. For our  present case, the entangling surface is just $S^1 \times S^1$, of which the Euler number is zero, hence only  the B-type anomaly would be singled out. It is indicated that there is no universal C-theorem for B-type anomaly, although in some theories one do have $C(UV) > C(IR)$ \cite{Anselmi:1997ys,Komargodski:2011vj}. We will then check the RG flow behavior of the renormalized entanglement entropy in the following.

\begin{figure}[t]
\begin{center}
\includegraphics[scale=0.8]{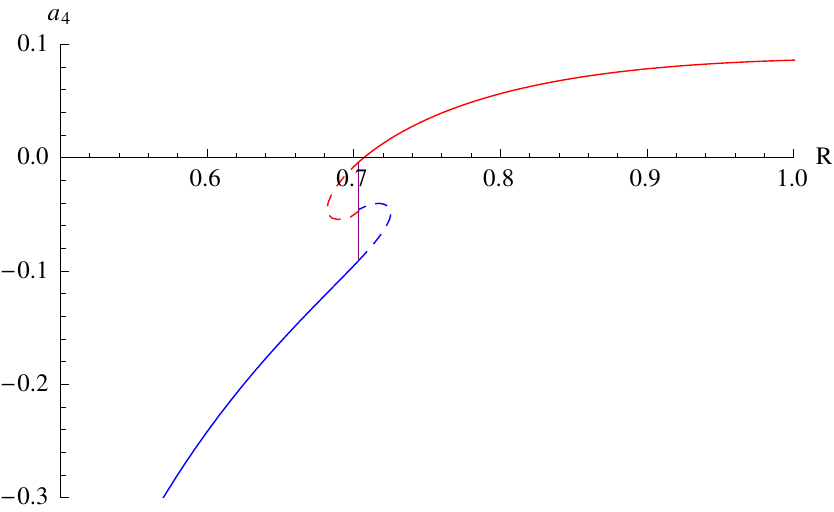}~~~~~
\includegraphics[scale=0.8]{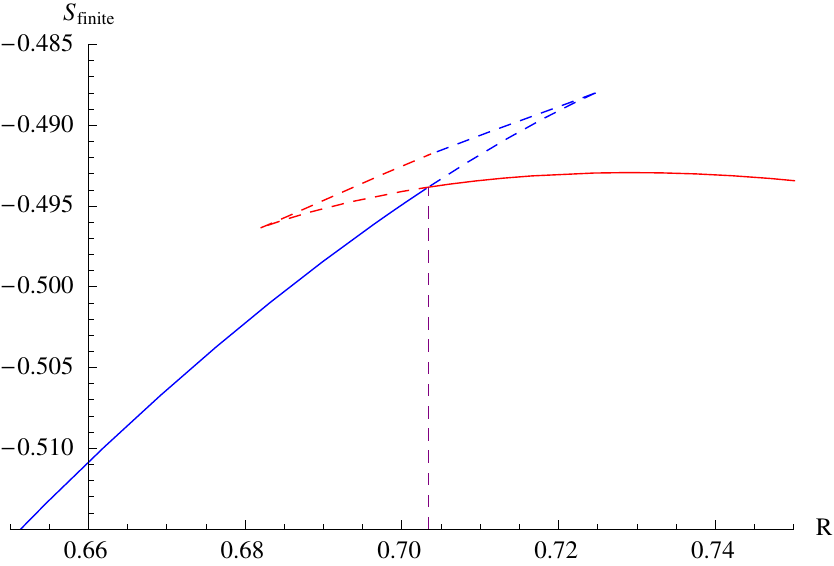}
\end{center}
\caption{Left: The $a_4(R)$ for AdS$_5$ soliton with $f(z)=1-z^4$.
Right: Finite part of on-shell action $S_{\rm finite}$ for the solutions around the critical point.}
\label{figa4AdS5soliton}
\end{figure}

  We numerically solve  $a_4(R)$  for the AdS soliton with $f(z)=1-z^4$ and the result is shown in the left plot of Fig.\,\ref{figa4AdS5soliton},
in which the blue and red curves denote contributions from disk and cylinder topologies, respectively.
The $a_4(R)$ is not single-valued near the phase transition between disk and cylinder topology. Since we have no other criterion for picking out a preferred
value of $a_4(R)$, to remove the additional branches we have to compare the on-shell actions of the solutions with both disk and cylinder topologies
around the critical point. Solutions with the smallest on-shell actions are chosen to be the dominant phase.

To determine the dominant topology, we introduce $S_{\rm finite}$ denoting the finite part of the on-shell action. For AdS$_5$ solitons, this is obtained by subtracting the divergent part $S_{\rm div}^{(4)} \sim \frac{R}{2 \epsilon^2} + \frac{1}{8R}\log{\frac{\epsilon}{R}}$ numerically from the total on shell action. The $S_{\rm finite}$ is different from the $S_{\rm UV-ind}$ defined in (\ref{RGAdS5}) in the sense that the UV cutoff-dependent terms not being removed. In fact,  $S_{\rm finite}$ is related to $S_{\rm UV-ind}$ via $S_{\rm UV-ind} = g_4(R\partial_R) S_{\rm finite}$. On the other hand, the $S_{\rm finite}$ could be used to determine the dominant phase, since it contains the total information of the entanglement entropy, up to a divergent part which is the same at every value of $R$ for different branches.

  The numerical results of the $S_{\rm finite}$ are shown in the right plot of
Fig.\,\ref{figa4AdS5soliton} \footnote{
  Note that the $S_{\rm finite}$ in Fig.\,\ref{figa4AdS5soliton} is negative such that it cannot be directly interpreted as the entanglement entropy or the number of degrees of freedom. Instead, $S_{\rm UV-ind}$ is positive at least in the UV regime as guaranteed by the large initial UV value ${1\over {8 R}}$ for small $R$, thus it could be thought as the number of degrees of freedom.} , which indicates that for $R< 0.703\,$ the disk topology dominates, while for $R>0.703$ the cylinder topology dominates. Hence the additional
branches of $a_4(R)$ in the left plot of Fig.\,\ref{figa4AdS5soliton} are removed. The discontinuous jump indicates a quantum phase transition. In fact, we can use the renormalized entanglement entropy $S_{\rm UV-ind}$ as an order parameter to characterize this quantum phase transition.

\begin{figure}[t]
\begin{center}
\includegraphics[scale=0.82]{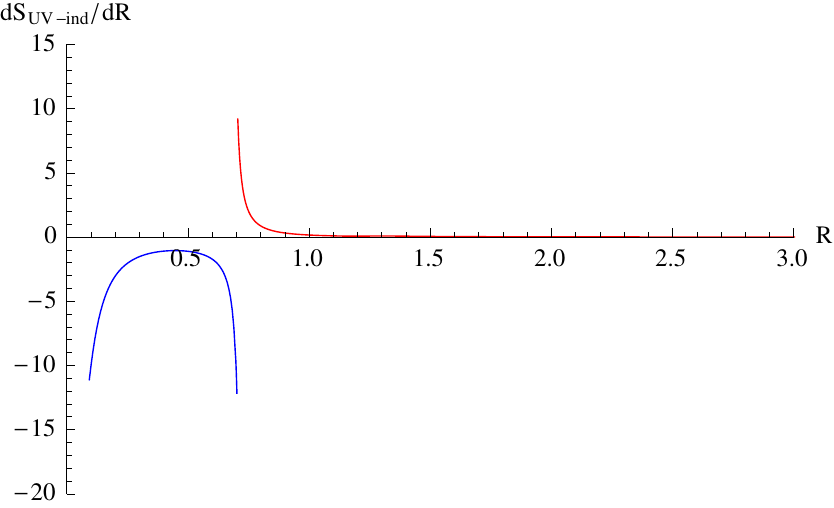}~~~~~
\includegraphics[scale=0.8]{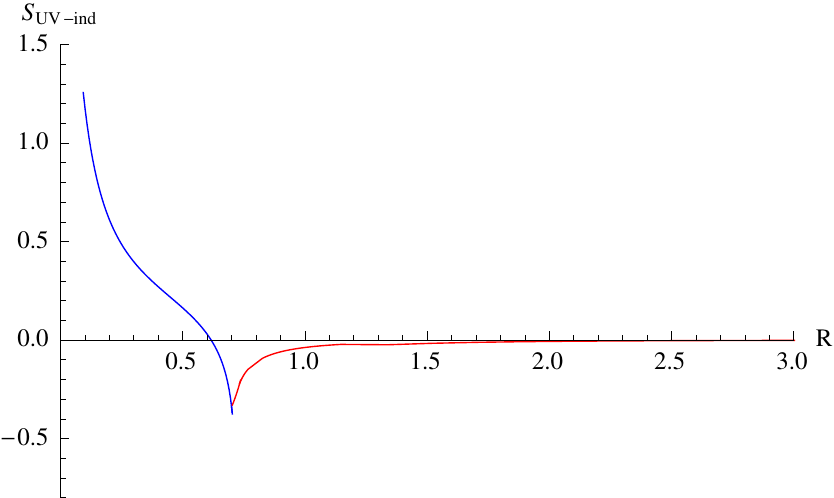}
\end{center}
\caption{Left: The ${dS^{(4)}_{\rm UV-ind}\over dR}$ for AdS$_5$ soliton
with $f(z)=1-z^4$. Right: The corresponding $S^{(4)}_{\rm UV-ind}$. } \label{figRGAdS5soliton}
\end{figure}

By using \eq{RGAdS5} and the results in  Fig.\,\ref{figa4AdS5soliton} we numerically calculate the RG flow of the renormalized entanglement entropy, ${dS^{(4)}_{\rm UV-ind}\over dR}$, which is shown in the left plot of Fig.\,\ref{figRGAdS5soliton}. We find there is a sharp jump around the critical point,  indicating the quantum phase transition from the deconfining phase in the UV regime to the confining phase in the IR regime. The C-theorem holds in the UV regime, which is expected since we define our subtraction scheme in the UV limit. However, the $S^{(4)}_{\rm UV-ind}$ becomes sharply increasing away from the critical point, indicating an increase of the number of degrees of freedom. This seems at odds with the C-theorem, however, since our choice of entangling surface singles out B-type anomaly \cite{Schwimmer:2008yh} and there is evidence that any combination that involves B-type anomaly dose not satisfy a-theorem in 4D \cite{Schwimmer:2008yh, Liu:2012ee, Komargodski:2011vj}, there should be no conflict with C-theorem.  
On the other hand, the renormalized entanglement entropy remains almost constant in the IR regime, which is consistent with expectation for the confining phase or the IR mean field state of a gapped systems, i.e., almost all the degrees of freedom are gapped out and the ground state is a trivial product state.

As a check of consistency, we also calculated $S^{(4)}_{\rm UV-ind}$ directly through $g_4(R\partial_R) S_{\rm finite}$ and the result is shown in the right plot of Fig.\,\ref{figRGAdS5soliton}. It is hard to tell whether the quantum phase transition is of first-order or second-order because of the numerical error. The fact that $S^{(4)}_{\rm UV-ind}$ is neither monotonic nor positive-definite is similar to the result of GPPZ flow obtained in \cite{Liu:2012ee}.

  We now consider  the special cases of AdS solitons, the extremal AdS solitons. For such cases, there are only solutions with disk topology. Technically, this fact could be realized from the IR expansion. Suppose that there exist solutions with cylinder topology, which end on $z=z_0$ at $r_0=r(z_0)$. We could expand the solution $r(z)$ around $r=r_0$ as following:
\be
r(z) = r_0 + d_1 (z_0 - z) + d_2 (z_0 - z)^2 + d_3 (z_0 - z)^3 + \cdot\cdot\cdot\,.
\ee
For non-extremal AdS soliton, we could work out the coefficients $d_1, \, d_2, \, d_3, \,\cdot\cdot\cdot$ order by order from the expansion of the equation of motion; however for extremal AdS soliton, one find that the coefficients $d_1, \, d_2, \, d_3, \,\cdot\cdot\cdot$ turn out to be infinity, which indicates that $z'(r_0)$ tends to zero. This means that one can never reach the boundary from $z=z_0$, that is, solutions with cylinder topology do not exist.

  On the other hand, the difference from the non-extremal case is that the proper size of the compact circle becomes infinite for the extremal one, this means that all the associated Kaluza-Klein (KK) modes become massless. That is, the dual field theory is a gapless system, and the IR behavior of the disk topology solution reflects this fact. More discussions on this will be given in section \ref{MERAsec}.

\begin{figure}[h]
\begin{center}
\includegraphics[scale=0.80]{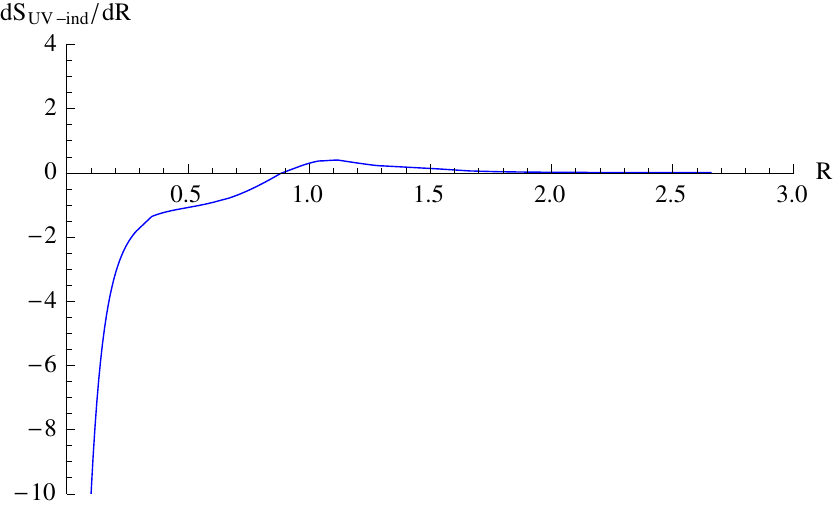}~~~~~
\includegraphics[scale=0.80]{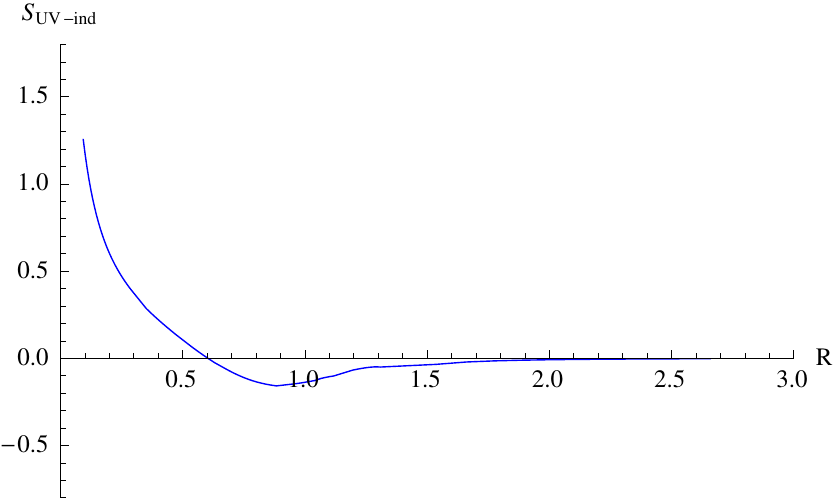}
\end{center}
\caption{Left: The ${dS^{(4)}_{\rm UV-ind}\over dR}$ for extremal charged AdS$_5$ soliton with $f(z) = 1 - 3 z^4 + 2 z^6$. \,Right: The corresponding $S^{(4)}_{\rm UV-ind}$. } \label{figRGexAdS5soliton}
\end{figure}

  By adopting the same differential subtraction scheme as for the non-extremal case, our numerical results for ${d S^{(4)}_{\rm UV-ind}\over dR}$ as well as $S^{(4)}_{\rm UV-ind}$ for extremal charged AdS$_5$ soliton with $f(z) = 1 - 3 z^4 + 2 z^6$ are shown in Fig.\,\ref{figRGexAdS5soliton}. Using again the $S^{(4)}_{\rm UV-ind}$ as the order parameter, we conclude that there is no phase transition. However, the monotonicity and positive-definiteness are still violated for $R$ greater than some specific value, and the IR behavior is also similar to the one for the non-extremal case.

\subsubsection{AdS$_6$ soliton}

      Similarly, we now consider the $d=5$ AdS soliton.  The UV expansion of the solution $r(z)$ takes the following form
\be\label{rexpanAdS6} r(z)=R-{z^2\over 3R}+{2(c_1-k_1-k_2)z^3\over 9 R z_0}+a_4(R) z^4
+ a_5(R) z^5 + \mathcal{O}(z^6) \ee where
 \be
a_4(R)={-k_1^2-k_2^2-k_1 k_2 +c_1 (k_1+k_2) -c_1^2+c_2 \over 6 R
z_0^2}-{5\over 54 R^3}
\ee
but $a_5(R)$ cannot be determined from
the UV expansion and should be solved from the full equation of
motion.  From the above expansion, we obtain
\be\label{dAdRAdS6} {d A \over dR}=-5
R^2 a_5(R) +{ 2(c_1-k_1-k_2) \over 3 R z_0}+ B {2 R \over
3z_0^3}+\;\mbox{UV-divergent terms}\; +\mathcal{O}(\epsilon),
\ee
where the coefficient $B$ depends only on the detailed form of the metric, i.e.,
\be
B=c_1^3-2c_1 c_2
+c_3-c_1^2k_1+c_2k_1+c_1k_1^2-k_1^3-c_1^2k_2+c_2k_2+c_1k_1k_2-k_1^2
k_2+c_1 k_2^2-k_1k_2^2-k_2^3.
\ee
The UV-divergent terms
take the form
\be
{2 R \over 3 \epsilon^3}.
\ee
Note that there is
no log divergent term as expected for $d={\rm odd}$ case. It seems a bit miraculous that there is also no ${\cal O}{(1/ \epsilon)}$ term in \eq{dAdRAdS6}, however there is such a term if we integrate \eq{dAdRAdS6} over $R$. To see this,
we substitute (\ref{rexpanAdS6}) into the action (\ref{Action}), expand the
integrand in series of $z$ and then integrate, we will find an additional divergent term
$-\frac{4}{9\epsilon}$. Since it is independent of $R$, we could not find
it in (\ref{dAdRAdS6}).
The $R$ scaling behaviors of the
UV-dependent terms are also different from the CFT case.

\begin{figure}[h]
\begin{center}
\includegraphics[scale=0.80]{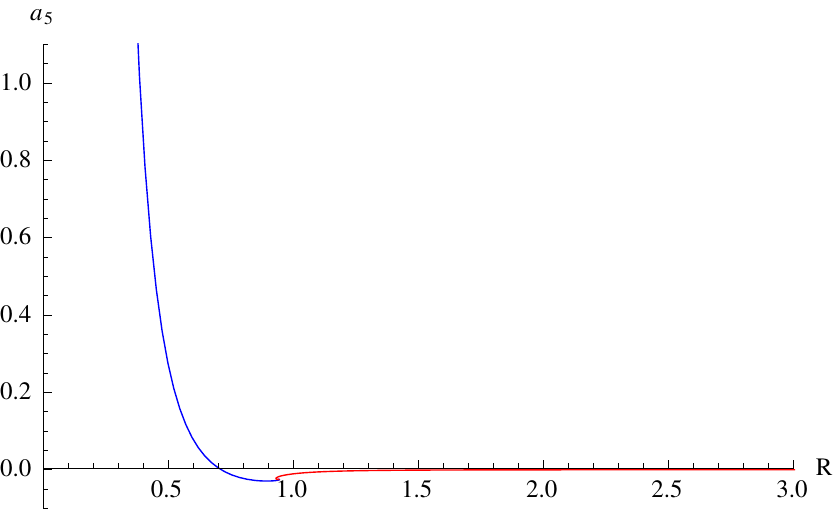}~~~~~
\includegraphics[scale=0.80]{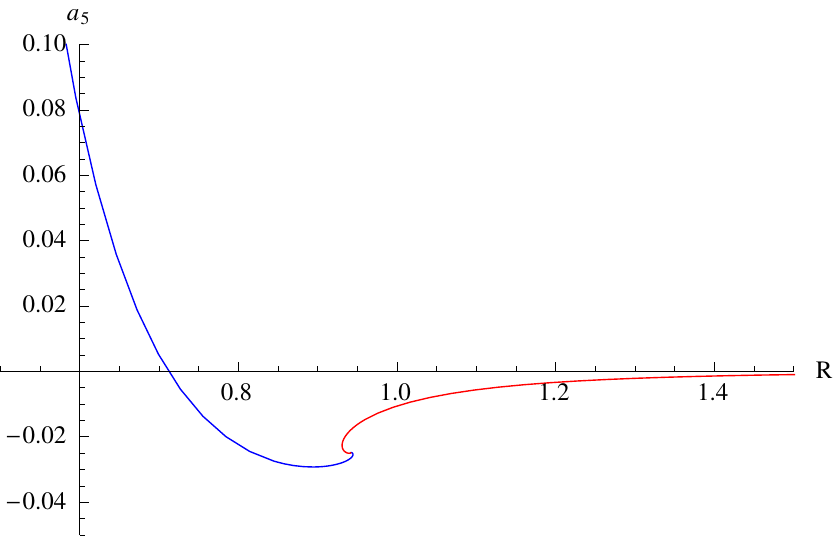}\\~\\
\includegraphics[scale=0.80]{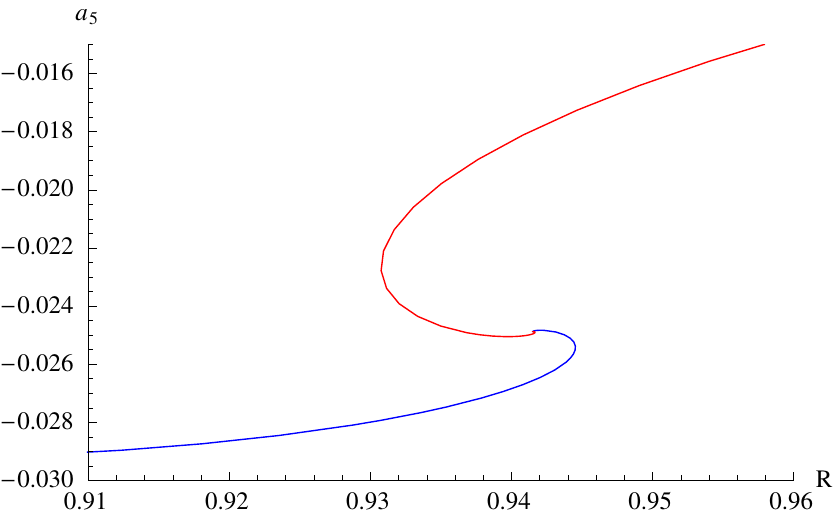}~~~~~
\includegraphics[scale=0.80]{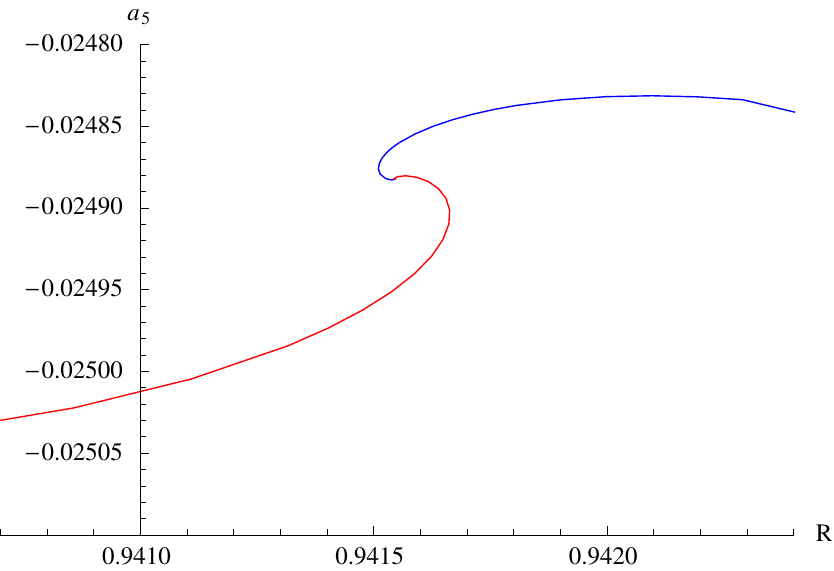}
\end{center}
\caption{The $a_5(R)$ on different scales for AdS$_6$ soliton with $f(z) = 1 - z^3\,$: fractal
vortex structure around the critical point. } \label{figa5fractal}
\end{figure}

We numerically solve $a_5(R)$  for the AdS$_6$ soliton with
$f(z)=1-z^3$ and the result is shown in Fig.\,\ref{figa5fractal}. Again the blue and red
curves denote contributions from disk and cylinder topologies, respectively. It is
interesting that near the critical point $R_c \sim 0.9415$, there seems to be a fractal
vortex structure, as is shown on different scales in Fig.\,\ref{figa5fractal}. This indicates
that $a_5(R)$ is multi-valued near the critical point.

\begin{figure}[h]
\begin{center}
\includegraphics[scale=0.80]{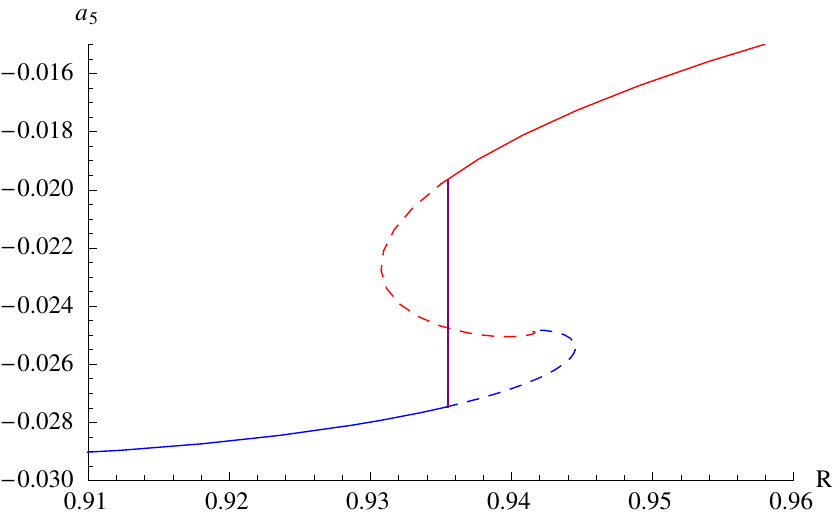}~~~~~
\includegraphics[scale=0.80]{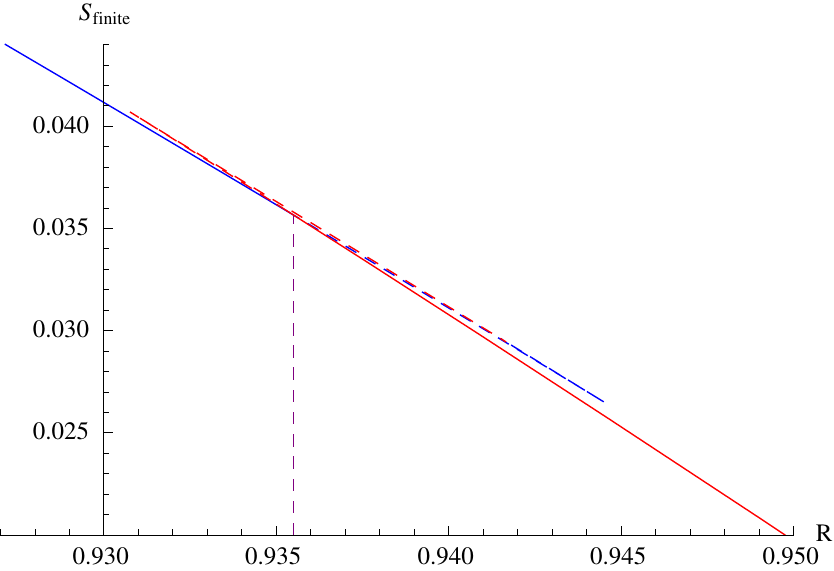}
\end{center}
\caption{Left: The $a_5(R)$ in detail for AdS$_6$ soliton with $f(z)=1-z^3$, with additional branches removed.
Right: Finite part of on-shell action for solutions around the critical point. } \label{figa5AdS6soliton}
\end{figure}

To determine the dominant phase, we
numerically calculated the finite part of the on-shell action with the divergence
$S_{\rm div} \sim \frac{R^2}{3 \epsilon^3} - \frac{4}{9 \epsilon}$ subtracted, and the
result is shown in the right plot of Fig.\,\ref{figa5AdS6soliton}. The situation is similar
to the right plot of Fig.\,\ref{figa4AdS5soliton} of the AdS$_5$ soliton case, though it is
a bit hard to distinguish the red and blue curves since they nearly coincide with each other.
From this plot we read the phase transition point $R = 0.9355$.
For $R < 0.9355$ the disk topology is dominant, while for
$R > 0.9355$ the cylinder topology is dominant. The $a_5(R)$ on the corresponding scale with
additional branches removed is shown in the left plot of Fig.\,\ref{figa5AdS6soliton}.
Note that the fractal vortex structure around $R_c \sim 0.9415$ is totally removed, hence it will not bring
additional phase transitions.

\begin{figure}[h]
\begin{center}
\includegraphics[scale=0.80]{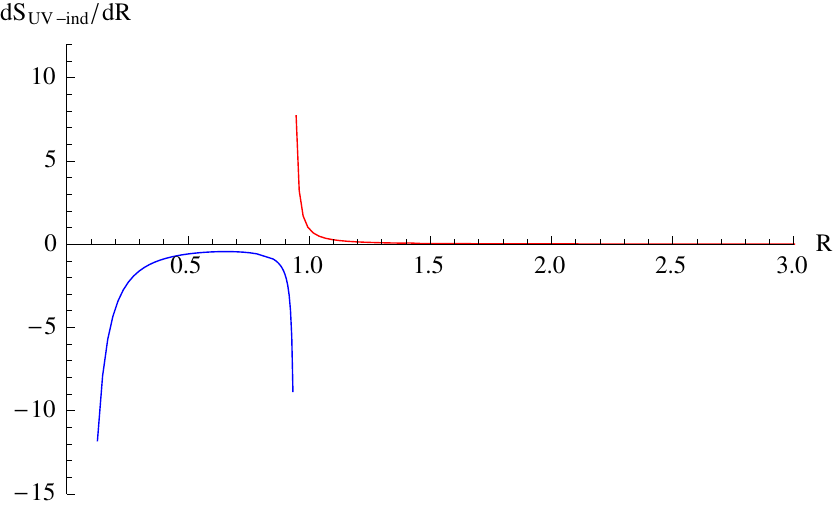}~~~~~
\includegraphics[scale=0.80]{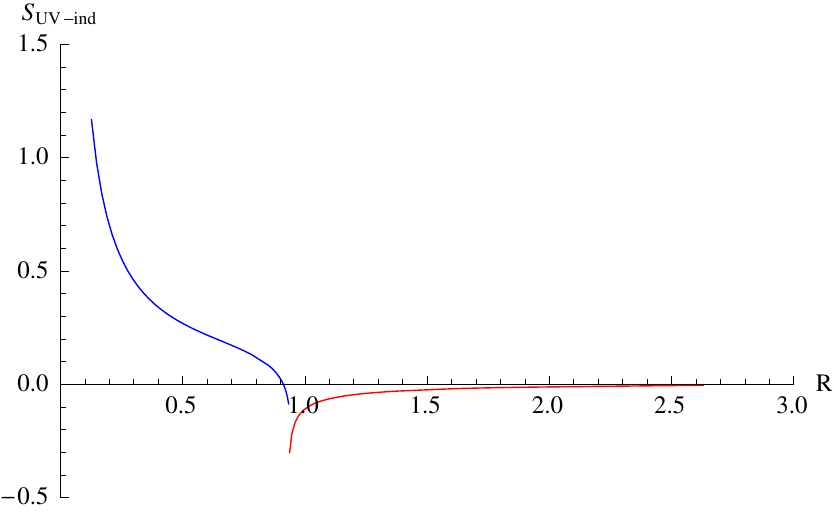}
\end{center}
\caption{Left: The ${d S^{(5)}_{\rm UV-ind} \over d R}$ for AdS$_6$ soliton with $f(z)=1-z^3$.
Right: The corresponding $S^{(5)}_{\rm UV-ind}$. }
\label{figRGAdS6soliton}
\end{figure}

The RG flow ${d S^{(5)}_{\rm UV-ind} \over d R}$ is calculated
straightforwardly by using \eq{diffsubs} for $d=5$ and is shown in the left plot of
Fig.\,\ref{figRGAdS6soliton}. The qualitative behavior is similar to the one for AdS$_5$ soliton.
It also indicates the occurrence of the deconfinement/confinement phase transition and
an increase of the number of degrees of freedom in the confining phase. The $S^{(5)}_{\rm UV-ind}$ is also
calculated through $g_5(R\partial_R) S_{\rm finite}$ and is shown in the right plot of
Fig.\,\ref{figRGAdS6soliton}, which is again neither monotonic nor positive-definite, indicating a
first-order phase transition.

\subsubsection{ AdS$_4$ soliton}

For completeness of the discussion on AdS solitons, we also give the results of the simplest $d=3$ AdS soliton.
The UV expansion of the solution $r(z)$ is simply
\be\label{rexpanAdS4}
r(z) = R + a_3(R) z^3 + {\cal O}(z^4)\,,
\ee
where $a_3(R)$ encodes the IR information and should be solved from the full equation of motion. From the above
expansion we obtain
\be\label{dAdRAdS4}
\frac{dA}{dR} = -3 \,a_3 (R) + {\cal O} (\epsilon)\,.
\ee
There is no UV-dependent divergent term in (\ref{dAdRAdS4}), but substituting (\ref{rexpanAdS4}) into the action (\ref{Action}) yields the divergent term
$1/\epsilon$. It is independent of $R$, hence does not appear in (\ref{dAdRAdS4}).

For $d=3$ AdS soliton, the cylinder solution is trivially $r(z) = R$,
as could be seen from the equation of motion (\ref{EOM}). As we will
see, it is the dominated topology for large $R$. From the action
(\ref{Action}) we obtain the on-shell action $S = {1 \over \epsilon}
- {1 \over z_0}$. For this case we have $a_3(R) =0 $ and $S_{\rm
finite} =  - {1 \over z_0}$.

\begin{figure}[h]
\begin{center}
\includegraphics[scale=0.80]{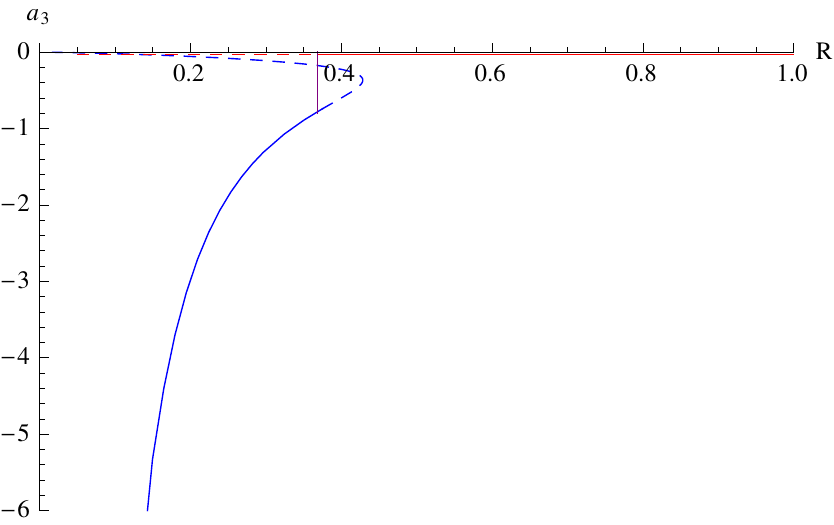}~~~~~
\includegraphics[scale=0.80]{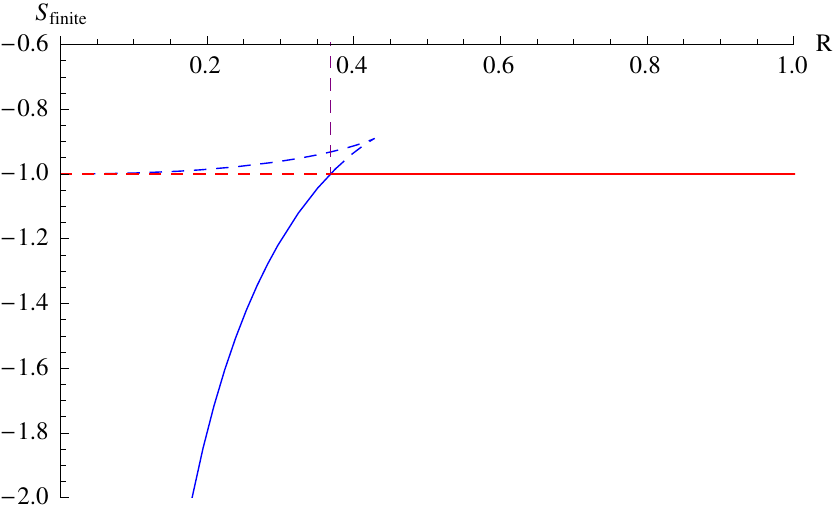}
\end{center}
\caption{Left: The $a_3(R)$ for AdS$_4$ soliton with $f(z)=1-z^5$.
Right: Finite part of on-shell action. } \label{figa3AdS4soliton}
\end{figure}

\begin{figure}[h]
\begin{center}
\includegraphics[scale=0.80]{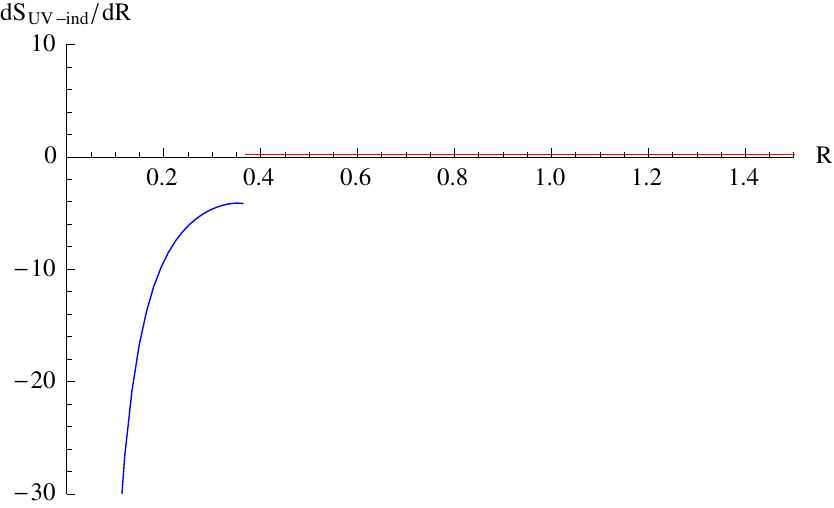}~~~~~
\includegraphics[scale=0.80]{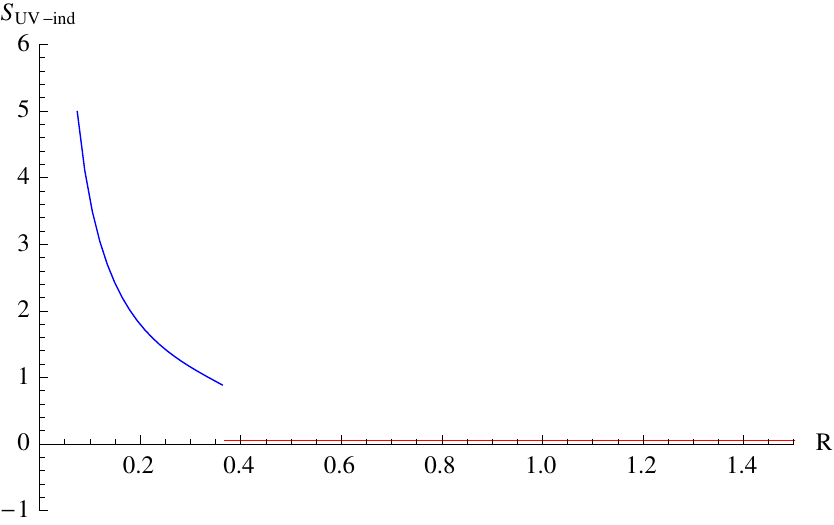}
\end{center}
\caption{Left: The ${d S^{(3)}_{\rm UV-ind} \over d R}$ for AdS$_4$ soliton with $f(z)=1-z^5$.
Right: The corresponding $S^{(3)}_{\rm UV-ind}$.} \label{figRGAdS4soliton}
\end{figure}

 We now concentrate on the disk solution which dominates over cylinder one at small $R$.
 The $a_3(R)$ and $S_{\rm finite}$ for AdS$_4$ soliton with $f(z)= 1-z^5$ is calculated numerically
and plotted in Fig.\,\ref{figa3AdS4soliton}, respectively. There is a phase transition at
$R = 0.3686$. For $R < 0.3686 $ the disk topology dominates, while for $R > 0.3686$ the
cylinder topology is dominant.

   Based on the above result, we now apply the differential scheme \eq{diffsubs} with $d=3$ to obtain
${d S^{(3)}_{\rm UV-ind} \over d R}$ for AdS$_4$ soliton with $f(z)=1-z^5$, and employ $g_3(R\partial_R) S_{\rm finite}$
to get the corresponding $S^{(3)}_{\rm UV-ind}$. Both of the results are shown in Fig.\ \ref{figRGAdS4soliton}.   There is no violation of C-theorem even there is a first-order phase transition at $R = 0.3686$. The renormalized entanglement entropy becomes constant after the quantum critical point. This is consistent with the expectation for the entanglement entropy of a $(1+1)$-dimensional gapped system.

\subsection{Extracting the topological entanglement entropy}

According to the study of the strongly coupled condensed matter systems,
the entanglement entropy contains both the short-range and the long-range ones\cite{SRE1,SRE2,SRE3,SRE4,SRE5}. The short-range
entanglement is responsible for the area law nature of the
entanglement entropy which measures the number of the entangled pairs with one particle inside the chosen region and the other one outside. On the other hand, the long-range entanglement  is a constant topological invariant, which is
independent of both the UV and IR scales, and should be associated
with existence of the topological order.  Especially, there are some
exactly solvable model with topological order in (2+1)-dimensions, and their entanglement entropies have the structure
\cite{Levin:2006zz,Kitaev:2005dm}
\be S=\alpha R -\gamma
\ee
where $\alpha$ and $\gamma$ are some constants. A nonzero $\gamma$
encodes the quantum dimensions of the anyonic excitations in the
topological ordered phase, and is called the topological entanglement
entropy.  See \cite{Grover} for the discussion of the topological entanglement entropy
for the higher dimensional theory, which  again should be a constant
piece in the entanglement entropy.

Since the topological entanglement entropy should be independent of the UV and IR scales, it should be encoded in the UV-independent piece. Note that the differential operator for AdS solitons (\ref{gsubs}) indicates
\be\label{gsubs2}
S_{\rm UV-ind} =\begin{cases} \frac{1}{(d-2)!!} \,R \frac{d}{d R} (R \frac{d}{d R} - 2) \cdots
(R \frac{d}{d R} - (d-3)) \,S, &
\qquad {\rm d \; odd}, \cr
\frac{1}{(d-2)!!} \,(R \frac{d}{d R} + 1) ( R \frac{d}{d R}-1) \cdots (R \frac{d}{d R} - (d-3))\, S, &
\qquad {\rm d \; even},
\end{cases}
\ee
hence the topological term would survive the differential operator in even dimensions, while in odd dimensions we could never observe such term in $S_{\rm UV-ind}$. Let's focus on AdS solitons in even dimension below.

To obtain $S_{\rm UV-ind}^{(d)}$ by integrating ${dS_{\rm UV-ind}^{(d)}\over dR}$ over $R$, one will get an integration constant. However, this constant is not relevant for topological order since it can be fixed by the UV part of the UV-independent piece, namely, $S_{\rm UV-ind}^{(d)}(R=0)$.  To look for the topological entanglement entropy encoding long-range entanglement, one instead should look for the IR behavior of the UV-independent piece. More precisely, one should extract the constant piece in the large $R$ expansion of $S^{(d)}_{\rm UV-ind}$. This piece will be independent of both UV and IR scales and should encode topological order.

 To avoid the numerical uncertainty, we here introduce an analytic way to extract the topological piece of entanglement entropy.
 The method is to consider the large $R$ expansion of both action \eq{Action} and the associated equations of motion, and then order by order solve $r_i$'s which are the coefficient functions in
\be
r(z)=r_0(z) R + r_1(z) + {r_2(z) \over R} + \mathcal{O}({1\over
R^2})
\ee
where $r_i$'s satisfy the UV boundary condition $r_0(0)=1$
and $r_{i\ne 0}(0)=0$ so that $r(0)=R$. It is easy to see that
$r_0(z)$ cannot be nontrivial from the leading order of equations of
motion. Thus we set $r_0(z)=1$.

   To be specific, we consider $d=4$ case. The action \eq{Action} in the large $R$ expansion is
\be\label{Action1}
A=\int_{\epsilon}^{z_m} dz \left( {\sqrt{1+f \dot{r}_1^2} \over z^3} R + {r_1 \sqrt{1+f \dot{r}_1^2} \over z^3}+{f \dot{r}_1 \dot{r}_2 \over z^3 \sqrt{1+f \dot{r}_1^2} }+\mathcal{O}({1\over R}) \right),
\ee
and the equation of motion in the large $R$ expansion yields
\be\label{r1eq}
0=R \partial_z({f \dot{r}_1  \over z^3 \sqrt{1+f \dot{r}_1^2}}) + \mathcal{O}({1\over R^0}).
\ee
Solving \eq{r1eq} with the boundary condition $r_1(0)=0$ yields $r_1(z)=0$ by using the fact that $f(0)=1$ and $\dot{r}_1(0)$ is finite such that $\dot{r}_1(0)=0$. This implies that the $R$-independent term in \eq{Action1} is zero irrespective of the topology of the holographic entangling hypersurface because we only use the UV geometry to yield $r_1(z)=0$. From (\ref{gsubs2}), we conclude that the topological entanglement entropy is zero for both extremal and non-extremal AdS$_5$ solitons.

\section{Considerations for the AdS$_4$ black holes}\label{AdSBHsec}

   We now consider another setting by turning on the temperature and chemical potential for the dual CFT. This is just to consider the AdS black hole with the following metric \cite{Hartnoll:2011fn} (to be specific we consider the AdS$_4$ planar black hole)
\be\label{BHmetric}
ds^2={L^2_{\rm AdS}\over z^2}\left(-f(z) dt^2 + {dz^2 \over f(z)}+ dr^2 + r^2 d\phi^2\right)
\ee
with
\be
f(z)=1-(1+{z_+^2 \mu^2 \over 2 \gamma^2})({z\over z_+})^3 + {z_+^2 \mu^2 \over 2 \gamma^2}({z\over z_+})^4,
\ee
where $\mu$ is the chemical potential for the dual CFT and the parameter $\gamma^2={e^2 L^2_{AdS} \over \kappa^2}$ is the dimensionless ratio of the Newtonian and Maxwell couplings. Moreover, the temperature $T$ of the black hole or the dual CFT is related to the position of horizon $z_+$ and chemical potential $\mu$ by
\be
T={1\over 4\pi z_+}(3-{z_+^2 \mu^2 \over 2 \gamma^2}).
\ee
The extremal black hole has $T=0$ by choosing ${z_+^2 \mu^2 \over 2 \gamma^2}=3$.

The thermal entropy density of the dual CFT is given by the Bekenstein-Hawking area law,
\be
s_{\rm thermal}={2\pi\over  \kappa^2} {A_h \over V_2} ={2\pi L^2_{\rm AdS} \over  \kappa^2 z_+^2}
\ee
where $V_2$ is the field theory volume and $A_h$ is the event horizon area.

  Now consider the holographic entanglement entropy in background \eq{BHmetric}. The entangling surface is defined by $z=0,\, r=R\,$ and $0\le \phi \le 2\pi$\,. The area of  the minimal surface is determined by the action
\be\label{Abh1}
A=\int \sqrt{\det g_{\rm ind}}= \int_{\epsilon}^{z_m} dz \; {r \over z^2} \sqrt{{1\over f}+ \dot{r}^2}:= \int_{\epsilon}^{z_m} dz \; \mathcal{L}.
\ee
From \eq{Abh1} we obtain
\be
\Pi={\partial \mathcal{L} \over \partial \dot{r}}={r \dot{r} \over z^2\sqrt{{1\over f}+\dot{r}^2}},\qquad \mathcal{H}=\Pi \dot{r}-\mathcal{L}=-{r\over z^2 \sqrt{f(1+f \dot{r}^2)}},
\ee
which appear in \eq{aRG}.

For the AdS black holes, one may also expect that there are minimal surfaces of either disk or cylinder topologies. Since the horizon is just the coordinate singularity, one might expect that the minimal surfaces of ``cylinder topology'' would extend into the region inside the horizon and have turning points there, that is, they are in fact of disk topology with the tip shadowed behind the horizon. However, it was pointed out in \cite{Hubeny:2012ry} that the minimal surfaces could not extend all the way to the horizon, hence they cannot penetrate the horizon. We therefore just focus on the solutions with disk topology.

   Solving  the equation of motion for the minimal surface in the UV expansion, we have
\be r(z)=R-{z^2\over 2R}+a_3(R) z^3 +\mathcal{O}(z^4) \ee where the
higher order terms are not relevant for UV-independent piece of the
entanglement entropy, and again $a_3(R)$ should be obtained by
solving the full range of the equation of motion, and depend on the
IR behavior of the minimal surface. Using \eq{aRG}, we have
\be
{d A \over d R}={1\over \epsilon} - 3 R a_3(R) +
\mathcal{O}(\epsilon^2).
\ee

  To extract the RG flow behavior of the renormalized entanglement entropy, we can apply the differential subtraction scheme given in \cite{Liu:2012ee}. By using the commutator relation $[\partial_R, R \partial_R]=\partial_R$, we obtain
\be\label{RGAdS4BH} {d S^{(3)\,{\rm
BH}}_{\rm UV-ind} \over d R}=  R \partial_R {d A\over dR}=
R \partial_R \left(- 3 R
a_3(R)\right).
\ee

\begin{figure}[h]
\begin{center}
\includegraphics[scale=0.8]{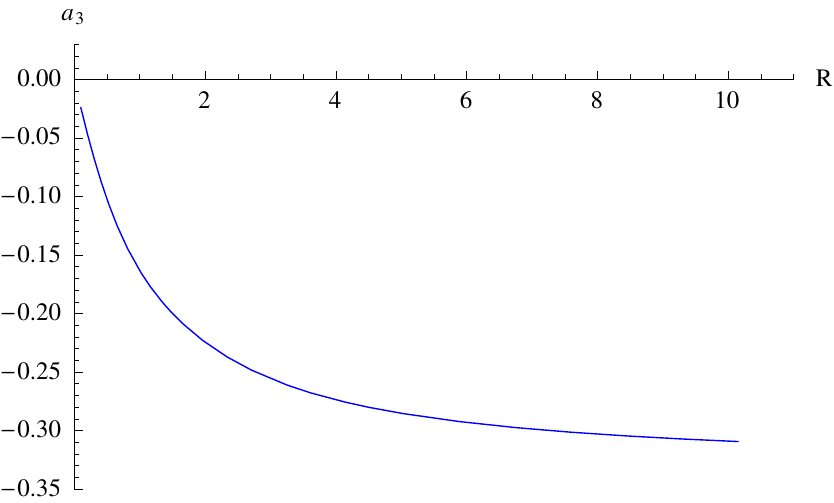}~~~~~
\includegraphics[scale=0.8]{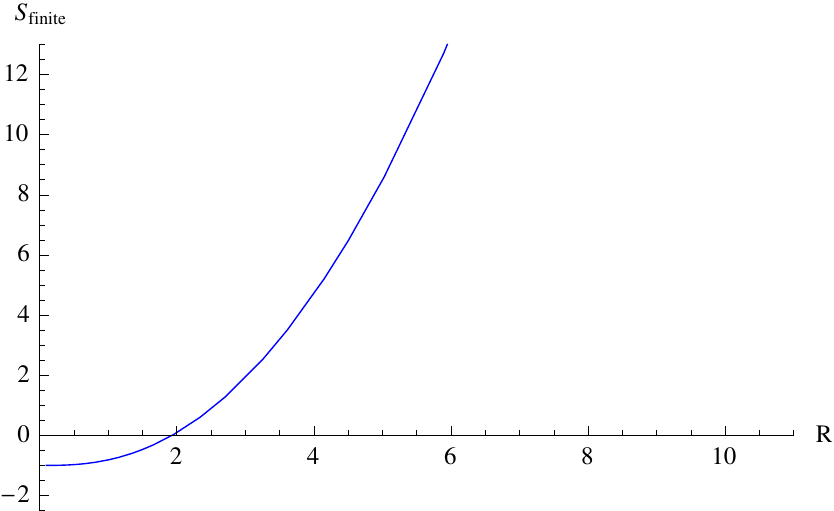}
\end{center}
\caption{Left: The $a_3(R)$ for non-extreme AdS$_4$ black hole with $f(z) = 1 - 2 z^3 + z^4$.  Right: Finite part of the on-shell actions. }
\label{figa3nonAdS4BH}
\end{figure}

  We first consider the case of non-extremal black hole.  We solve $a_3(R)$ numerically and the result is shown in the left plot of Fig.\,\ref{figa3nonAdS4BH}. The on-shell actions with divergent part $S_{\rm div}^{(3)\,BH}\sim R/\epsilon$ subtracted are shown in the right plot of Fig.\,\ref{figa3nonAdS4BH}.
Since there are only solutions with disk topology, there would be no phase transition along the RG flow.

\begin{figure}[h]
\begin{center}
\includegraphics[scale=1.00]{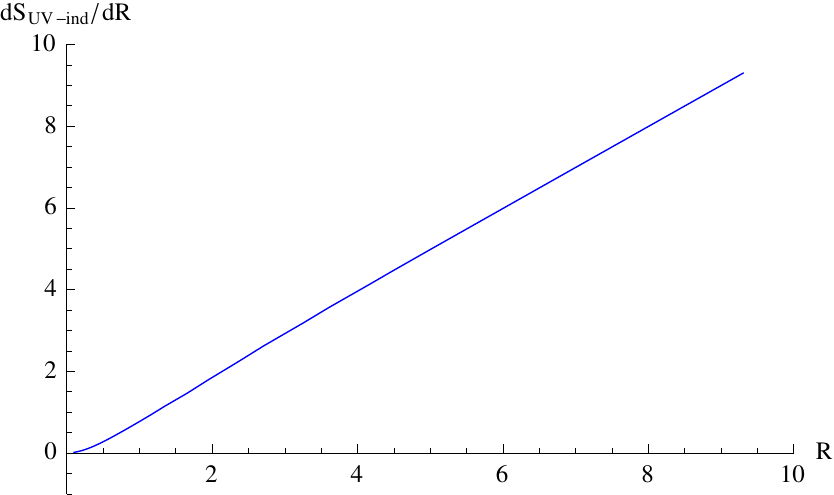}
\end{center}
\caption{The ${dS^{(3)\,{\rm BH}}_{\rm UV-ind}\over dR}$ for
non-extreme AdS$_4$ black hole with $f(z) = 1 - 2 z^3 + z^4$. }
\label{figRGnonAdS4BH}
\end{figure}

\begin{figure}[h]
\begin{center}
\includegraphics[scale=0.8]{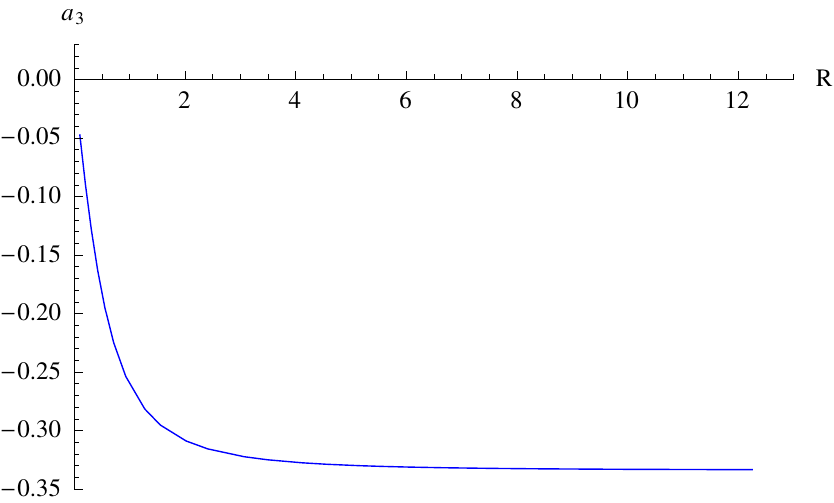}~~~~~
\includegraphics[scale=0.8]{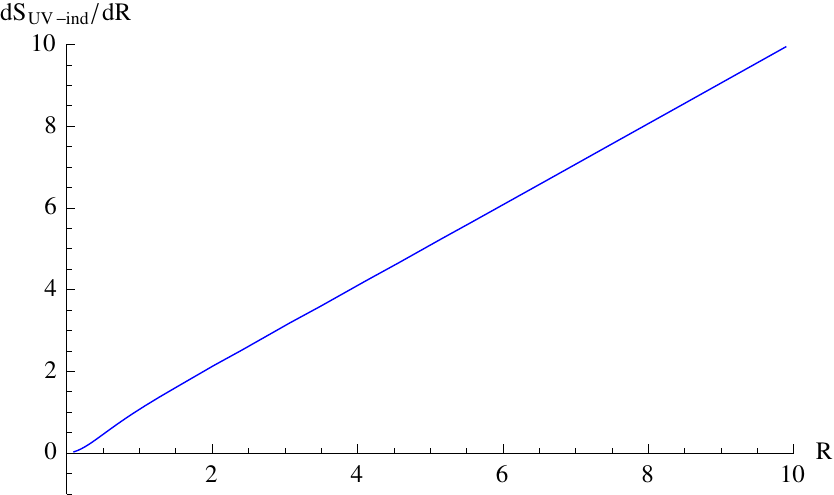}
\end{center}
\caption{Left: The $a_3(R)$ for extreme AdS$_4$ black hole with $f(z) =
1 - 4 z^3 + 3 z^4$. Right: The corresponding ${dS^{(3)\,{\rm BH}}_{\rm
UV-ind} \over dR}$. } \label{figa3exAdS4BH}
\end{figure}

  The RG flow of  the renormalized entanglement entropy is then followed from  (\ref{RGAdS4BH}) and the numerical result is shown in Fig.\,\ref{figRGnonAdS4BH}.  We see that ${dS^{(3)\,{\rm BH}}_{\rm UV-ind} \over dR}$  is always positive, implying that more and more states are thermally excited as we go to higher temperature regime, or equivalently lower energy scale.
 For large $R$ it is consistent with the linear running of the thermal entropy, i.e.,
\be
{d S^{(3)}_{\rm thermal} \over d R}= 2\pi R \; s_{\rm thermal}.
\ee
Hence there is no phase transition but a smooth crossover interpolating between the renormalized entanglement entropy for the ground state in the IR regime and the
thermal entropy in the UV regime. This fact supports the conjecture proposed in \cite{Swingle:2011mk} \footnote{Their conjecture refers to the finite part of the entanglement entropy, the $S_{\rm finite}$. Instead, we are considering the UV-independent piece, the $S_{\rm UV-ind}$. In this sense, our results yield a refined version of their conjecture. One main difference is that the area-law piece of $S_{\rm finite}$ in the UV regime is UV-ambiguous and cannot be included in $S_{\rm UV-ind}$. In contrast, the volume-law piece in the IR regime is UV-unambiguous.}.

    We also consider the case of  extremal black hole, its $a_3(R)$ and the RG flow of $S^{(3)}_{\rm UV-ind}$ are solved numerically and the results are  plotted in
Fig.\,\ref{figa3exAdS4BH}. Again we see the crossover from the IR regime to the UV one.

\section{Considerations for the AdS$_5$ solitons and black holes with Gauss-Bonnet correction}\label{GBsec}

   In this section we will consider the effect of the Gauss-Bonnet term to the refinement of the holographic entanglement entropy for both AdS$_5$ soliton and black hole. The bulk theory we consider is given by the action
\be
I=-{1\over 16 \pi G_N} \int d^5x \sqrt{g} \left( -{12\over L^2 } +\mathcal{R} +{\lambda_{GB} L^2  \over 2} \mathcal{L}_{GB} \right)
\ee
where $\lambda_{GB}$ is the coupling constant for the Gauss-Bonnet term with the Lagrangian
\be
\mathcal{L}_{GB}=\mathcal{R}_{\mu\nu\rho\sigma}\mathcal{R}^{\mu\nu\rho\sigma}-4 \mathcal{R}_{\mu\nu}\mathcal{R}^{\mu\nu}+\mathcal{R}^2.
\ee
In the dual theory, the higher curvature terms correspond to some perturbation in the sub-leading order of inverse 't Hooft coupling.

  The Gauss-Bonnet coupling $\lambda_{GB}$ should be in the interval $[0,1/4]$ for the metric to be well-defined outside the horizon of the AdS black hole or the IR end-point of the AdS soliton.  Moreover, for the black hole in $(4+1)$-dimensional AdS-Einstein-Gauss-Bonnet gravity theory,  it was shown in \cite{Brigante:2008gz,Buchel:2009tt} that the dual CFT will violate microcausality and render inconsistency when $\lambda_{GB}>{9\over 100}$.   We will then explore this effect to the refinement of entanglement entropy by studying ${d S_{\rm UV-ind}\over dR}$  for various values of $0\le \lambda_{GB}\le 1/4$.  At the same time, we will check whether the Gauss-Bonnet term would affect the crossover from the UV regime to the volume law in the IR regime. On the other hand, for the boundary dual theory of the AdS soliton we will simply pick a specific value of $0\le \lambda_{GB}\le 1/4$ in the following discussion.

  We will now first consider the RG flow of the renormalized entanglement entropy for the Gauss-Bonnet corrected AdS soliton, and then for the corrected black hole.

\subsection{Renormalized entanglement entropy for the Gauss-Bonnet corrected soliton}

The AdS soliton solution in $(4+1)$-dimensional AdS-Einstein-Gauss-Bonnet gravity theory is given by the metric \footnote{In \cite{Ogawa:2011fw}, the UV divergence structure of the holographic entanglement entropy of this metric for the stripe region has been studied. They also studied the entropic phase transition by varying $\lambda_{GB}$.}  \cite{Cai:2001dz}
\be\label{metricGB}
ds^2 = L^2 \left( \frac{dz^2}{z^2 f(z)}
+ \frac{L^2}{L^2_{\rm AdS}} \frac{dx_\mu dx^\mu}{z^2} + f(z)\frac{d\theta^2}{z^2}  \right)\,,\quad\quad \mu = 0,1,2
\ee
where
\bea\label{fzdef}
f(z) &=& \frac{1}{2\lambda_{GB}}\left(1 - \sqrt{1-4\lambda_{GB}\left\{1-\Big(\frac{z}{z_0}\Big)^4\right\}}\,\right)\,,\\ \label{f0def}
f_0 &=& \lim_{z\to\,0}f(z) = \frac{2}{1+\sqrt{1-4\lambda_{GB}}}\,,
\eea
\be
L_{\rm AdS} = \frac{L}{\sqrt{f_0}}\,, \qquad \mbox{and} \qquad
\theta\sim\theta + L_{\theta} \,, \qquad L_{\theta} = \pi  z_0\,.
\ee
For $\lambda_{GB} \to 0$, the metric (\ref{metricGB}) reduce to the AdS soliton part of (\ref{metric1}).  Note that $L$ is different from $L_{\rm AdS}$, and in the numerical study of this section we will set $L=1$ instead of $L_{\rm AdS}=1$.  From \eq{fzdef} and \eq{f0def} it is easy to see that $\lambda_{GB}$ should be in the interval $[0,1/4]$ so that the metric \eq{metricGB} has the well-defined Euclidean section for $0<z<z_0$.

\vspace*{4mm}

Consider a disk on the boundary with radius $R$, the induced metric of the minimal surface is given by
\be\label{indmetricGB}
ds^2_{\rm ind} = L^2 \left( \frac{1}{z^2}\left( \frac{L^2 }{L^2_{\rm AdS}}\, {\dot r}(z)^2 + \frac{1}{f(z)} \right)dz^2
+ \frac{L^2}{L^2_{\rm AdS}} \frac{r(z)^2}{z^2}\, d\phi^2 + \frac{f(z)}{z^2}\,d\theta^2  \right)\,,
\ee
where $r$ and $\phi$ are the radial and angular coordinates of the disk respectively. The minimal surface is determined by specifying $r(z)$.

\vspace*{4mm}

The holographic entanglement entropy in the Gauss-Bonnet gravity is
given by minimizing the functional
\cite{Fursaev:2006ih,Ogawa:2011fw} \be\label{heeGB}
A= \int_{\gamma_A}dx^3\sqrt{h}\,(1+\lambda_{GB} L^2
{\cal R}) + 2 \,\lambda_{GB}
L^2 \int_{\partial\gamma_A}dx^2\sqrt{h_b}\,{\cal K}\,, \ee
where $\cal R$ is the intrinsic curvature of the induced metric
$h$\,; $h_b$ is the induced metric on $\partial\gamma_A$ and $\cal
K$ is the trace of its extrinsic curvature.  The second term is the
so-called Gibbons-Hawking term. From (\ref{indmetricGB}) we obtain
\be \sqrt{h}\,(1+\lambda_{GB} L^2 {\cal R})= \frac{L^4\, r
}{L^2_{\rm AdS}\, z^3  } \sqrt{L^2_{\rm AdS} + L^2 f {\dot r}^2} +
\lambda_{GB} \frac{L^4 \left( 2 f\, r - z \dot{f} \,r- 2\, z f\,
{\dot r} + z^2 \dot{f}{\dot r} \right) } { z^3 \sqrt{L^2_{\rm AdS} +
L^2 f {\dot r}^2} }+\dot{q} \,, \ee where \be q(z) =
\lambda_{GB}\frac{  L^4 \left(4 f\, r - z \dot{f}\,r - 2 \,z f\,
{\dot r}\right)} {z^2 \sqrt{L^2_{\rm AdS} + L^2 f{\dot r}^2}}\,. \ee
Integrating the term $\dot{q}(z)$ on $\gamma_A$ gives rise to a
surface term which cancels the Gibbons-Hawking term in
(\ref{heeGB}). Therefore, the functional we need to minimize is
\bea\label{actionGB} A  &=&
\int^{z_m}_\epsilon dz \left( \frac{L^4\, r}{L^2_{\rm AdS}\, z^3  }
\sqrt{L^2_{\rm AdS}  + L^2 f {\dot r}^2} +  \lambda_{GB}\, \frac{L^4
\left( 2 f\, r - z \dot{f} \,r - 2\, z f\, {\dot r} + z^2
\dot{f}{\dot r}\right) } { z^3 \sqrt{L^2_{\rm AdS} + L^2 f {\dot
r}^2} } \right) \nn
\\
&:=&\int dz\; \mathcal{L}\,.
\eea
The equation of motion for (\ref{actionGB}) turns out to be
\begin{multline}\label{eomGB}
0 =
L^6 f^2 r ( -\,6 f + z \dot{f} ) {\dot r}^5
+ 2\, L^6_{\rm AdS} \,z\, (-1 + 2\,\lambda_{GB} f - 2 \,\lambda_{GB} z \dot{f} + \lambda_{GB} z^2  \ddot{f}) \\
 + L^4  L^2_{\rm AdS} \,f\, {\dot r}^2
\left[
\,z \,\dot{f}\, r\, (\,3 - \lambda_{GB}z \dot{f})\, {\dot r}
+ 4 \,\lambda_{GB} f^2 \left( -2\, z\, {\dot r}^2 + r (3\, {\dot r} + 2\, z\, {\ddot{r}}) \right) \right.\\
+ 2 \,f
\left(
z \,(-1 + 2 \,\lambda_{GB}\,z \dot{f}){\dot r}^2 + r \left( {\dot r} (-6 - 3 \,\lambda_{GB}\,z \dot{f} + \lambda_{GB} \,z^2  \ddot{f})\right. \right. \\
\left. \left. \left.+ \,z\, (1 - 2 \,\lambda_{GB}\,z \dot{f}) {\ddot r}   \right)\right) \right]
 + L^2 L^4_{\rm AdS}
\left[
\,z \dot{f}\, {\dot r} \left( 2\, r (1 +  \lambda_{GB}\,z \dot{f}) - 3 \,\lambda_{GB}\,z^2  \dot{f} {\dot r} \right)\right. \\
+ 4 \,\lambda_{GB} f^2 \left( r(z) (3 {\dot r} - z \,{\ddot r}) + z \,{\dot r} (-{\dot r} + 3\, z\, {\ddot r}) \right)
 + 2 f
\left(
z\, {\dot r} \left( {\dot r} (-2 + 3 \, \lambda_{GB} \,z \dot{f} +  \lambda_{GB}\,z^2 \ddot{f})\right.\right. \\
\left. - \,3 \,\lambda_{GB}\,z^2  \dot{f}) {\ddot r}  \right)
\left.\left. +~ r \left( {\dot r} (-3 - 6 \,\lambda_{GB}\,z \dot{f} + \lambda_{GB}\,z^2  \ddot{f})
+ z \,( 1 + \lambda_{GB} \,z \dot{f} ) {\ddot r} \right)\right) \right]   \,.
\end{multline}

%%%%%%%%%%%%%%%%%%%%%%%%%%%%%%
\subsubsection{Solutions of the minimal surfaces}
%%%%%%%%%%%%%%%%%%%%%%%%%%%%%%%%%%%%%%%%%
\begin{figure}[htbp]%[H]
\includegraphics[scale=1.00]{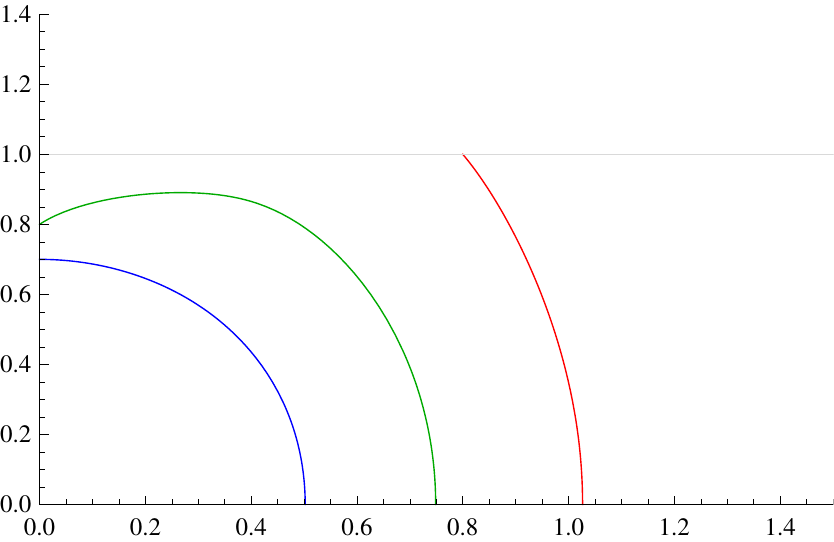}
\caption{Disk (blue), Cusp (Green) and  cylinder (red)
solutions of the minimal surface for AdS soliton with higher
derivative correction with $\lambda_{GB}=0.05$, $z_0=1$. Since there are
infinite number of cusp solutions with the same  $R$, we plot a
typical one.} \label{hisol}
\end{figure}
%%%%%%%%%%%%%%%%%%%%%%%%%%%%%%%%%%%%%%%%%%

 First, we consider the solution of (\ref{eomGB}) in the small $R$
 regime. In this regime, the solution has a disk topology as the
 blue line in Fig. \ref{hisol}. Near $r=0$ it can be expanded as
\begin{equation}
z(r)=z_m+z_2r^2+{\cal O}(r^4), \label{eqzm}
\end{equation}
where $z_m$ is defined as
\begin{equation}
z_m=z(r=0). \label{higherzm}
\end{equation}
By plugging (\ref{eqzm}) in (\ref{eomGB}), we find that the
coefficient $z_2$ satisfies the following quadratic equation,
\begin{equation}\label{z2abc}
az_2^2+bz_2+c=0,
\end{equation}
where
\begin{eqnarray}
a&=&-12z_0^4z_m^2\lambda_{GB}^2(-1-\gamma+2\lambda_{GB})[2z_m^4\lambda_{GB}+z_0^4(1-\xi+2\lambda_{GB}(\xi-2))],\\
b&=&-2z_0^4z_m(1+\gamma)\lambda_{GB}[2z_m^4\lambda_{GB}(-8+8\lambda_{GB}+3\xi)
\nn \\
&&+ \, z_0^4(-5+24\lambda_{GB}-16\lambda_{GB}^2+5\xi-14\lambda_{GB}\xi)],\\
c&=&8z_m^8\lambda_{GB}^2+z_0^4z_m^4\lambda_{GB}(13-20\lambda_{GB}-7\xi)+3z_0^8\left(1+4\lambda_{GB}^2-\xi+\lambda_{GB}(-5+3\xi)\right),
\end{eqnarray}
and for simplicity, we introduce  $\gamma$ and $\xi$ as
\begin{equation}
\gamma\equiv \sqrt{1-4\lambda_{GB}}, \label{higm}
\end{equation}
\begin{equation}
\xi\equiv
\sqrt{1-4\lambda_{GB}\left(1-\left(\frac{z_m}{z_0}\right)^4\right)}.
\end{equation}

%%%%%%%%%%%%%%%%%%%%%%%%%%%%%%%%%%%%%%%%
\begin{figure}[htbp]%[H]
\vspace{.3cm}
\includegraphics[scale=1.00]{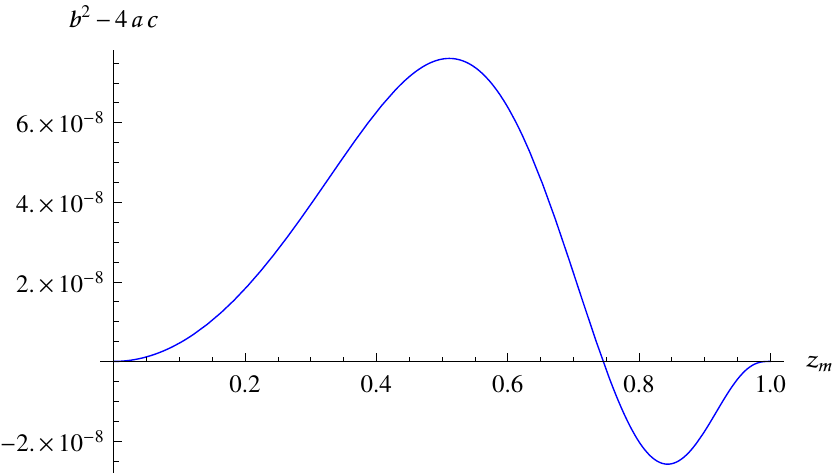}
\caption{Relation between $b^2-4ac$ and $z_m$ with $\lambda_{GB}=0.05$ and
$z_0=1$} \label{abcc}
\end{figure}
%%%%%%%%%%%%%%%%%%%%%%%%%%%%%%%%%%%%%%%%%%
The discriminant $D(z_m)\equiv b^2-4ac$ of (\ref{z2abc}) as a function of $z_m$ is shown in Fig. \ref{abcc}. There is $z_m=z_d$ which satisfies $D(z_d)=0$. Therefore, the solutions of disk topology exist only for
\be
0\le z_m\le z_d \hspace{5mm}\text{i.e.,}\hspace{5mm}0\le R\le
R_d\equiv R(z_d).
\ee
In the following we will  take $\lambda_{GB}=0.05$ and $z_0=1$, and in this case $R_d=0.528$.

Next, we consider the solution of (\ref{eomGB}) in the large $R$
 regime. In this regime, the solution has a cylinder topology as the
 red line in Fig. \ref{hisol}. Expand the solution near $z=z_0$ as
\begin{equation}
r(z)=r_0+r_1(z-z_0)+{\cal O}\left((z-z_0)^2\right),
\end{equation}
where
\begin{equation}
r_0\equiv r(z_0),
\end{equation}
and
\begin{equation}
r_1=\frac{-r_0\gamma^2+\gamma\sqrt{r_0^2\gamma^2-3z_0^2(1+\gamma)\lambda_{GB}(1+8\lambda_{GB})}}{12z_0\lambda_{GB}}.
\label{hir1}
\end{equation}
For
\begin{equation}
r_0=\frac{\sqrt{3z_0(1+\gamma)\lambda_{GB}(1+8\lambda_{GB})}}{\gamma}\equiv r_{\rm cyl},
\label{hir0}
\end{equation}
the expression inside the square root in (\ref{hir1}) becomes zero. Thus,
the solutions of cylinder topology exist only for
\begin{equation}
r_{\rm cyl}\le r_0  \hspace{5mm}\text{i.e.,}\hspace{5mm} R_{\rm cyl}\equiv
R(r_{\rm cyl})\le R.
\end{equation}
In the case of $\lambda_{GB}=0.05$ and $z_0=1$, $R_{\rm cyl}=0.963$.

For $R_d<R<R_{\rm cyl}$,  solutions of (\ref{eomGB}) have a cusp shape as shown in Fig. \ref{hisol}. For this solution, $z'(r=0)\neq 0$.
However, the cusp solutions for a fixed $R$ are not unique because we can adjust both $z(r=0)$ and $z'(r=0)$ to have the same $R$ at UV
\footnote{In fact for $R<R_d$ and $R>R_{\rm cyl}$ there are also cusp solutions, which were ignored because of their non-uniqueness.}. The absence of the smooth solution and the non-uniqueness of the cusp solutions suggests that there is no saddle point for prescription of \cite{Ryu:2006ef} in evaluating the holographic entanglement entropy in this regime of $R$.  This may suggest the need of some quantum version of prescription of \cite{Ryu:2006ef} to deal with such a case. Since we do not have such a prescription yet, in the following we will just skip discussion of the RG behavior for this regime.

\subsubsection{Renormalized entanglement entropy and its RG flow}

 Recall \eq{aRG} for the RG flow of the on-shell action,
\begin{equation}
\frac{dA}{dR}=-{\cal H}(z_m)\frac{dz_m}{dR}-\Pi(\epsilon)\frac{dr(\epsilon)}{dR},
\label{hids}
\end{equation}
where
\begin{eqnarray}
\Pi&=&\frac{\delta \mathcal{L}}{\delta\dot{r}}\hspace{142mm}\nonumber\\
&=& \frac{L^4\left(z^2\lambda_{GB} L_{AdS}^4\dot{f}+fL_{AdS}^2(-2z\lambda_{GB}
L_{AdS}^2+L^2r(1+z\lambda_{GB} \dot{f})\dot{r})+L^2f^2r\dot{r}(-2\lambda_{GB}
L_{AdS}^2+L^2\dot{r}^2)\right)}{z^3L_{AdS}^2(L_{AdS}^2+L^2f\dot{r}^2)^{3/2}}\nn\\
\end{eqnarray}
and
\begin{eqnarray}
{\cal H}&=&\Pi\dot{r}-{\cal L}\hspace{132mm}\nonumber\\
&=&-\frac{L^4\left(L^2z\lambda_{GB} f(-2f+z\dot{f})\dot{r}^3+r(L^2_{AdS}(1+2\lambda_{GB}
 f-z\lambda_{GB}\dot{f})+L^2f(1+4\lambda_{GB} f-2z
 \lambda_{GB}\dot{f})\dot{r}^2)\right)}{z^3(L^2_{AdS}+L^2f\dot{r}^2)^{3/2}}.\nn\\
\end{eqnarray}

 After simplification, the first term in (\ref{hids}) becomes
\begin{eqnarray}
{\cal
H}(z_m)\frac{dz_m}{dR}&=&\frac{L^3\lambda_{GB}\left(2f(z_m)-z_m\dot{f}(z_m)\right)}{z_m^2\sqrt{f(z_m)}}\frac{dz_m}{dR}, \text{\hspace{5mm}for
disk  topology},\\
{\cal H}(z_m)\frac{dz_m}{dR}&=&0, \text{\hspace{5mm}for cylinder
topology because }\hspace{3mm} \frac{dz_m}{dR}=\frac{dz_0}{dR}=0 .
\end{eqnarray}
Note that it is not zero for the disk topology, unlike the case with $\lambda_{GB}=0$.

The UV behavior of the solution $r(z)$ is
\begin{equation}
r(z)=R+a_2 z^2+a_4 z^4+b_4 z^4\log { z \over R }+\cdots, \label{hir}
\end{equation}
where
\begin{eqnarray}
a_2=-\frac{1+\gamma-2\lambda_{GB}}{8R},\qquad\qquad
b_4=\frac{(1+\gamma)\lambda_{GB}^3}{32R^3(1-\gamma-(3-\gamma)\lambda_{GB})}.
\end{eqnarray}
Again the coefficient $a_4$ cannot be determined from the UV expansion, and should be solved from the full equation of motion.

%%%%%%%%%%%%%%%%%%%%%%%%%%%%%%%%%%%%%%%%
\begin{figure}[b]%[H]
\vspace{.3cm}
\includegraphics[scale=1.00]{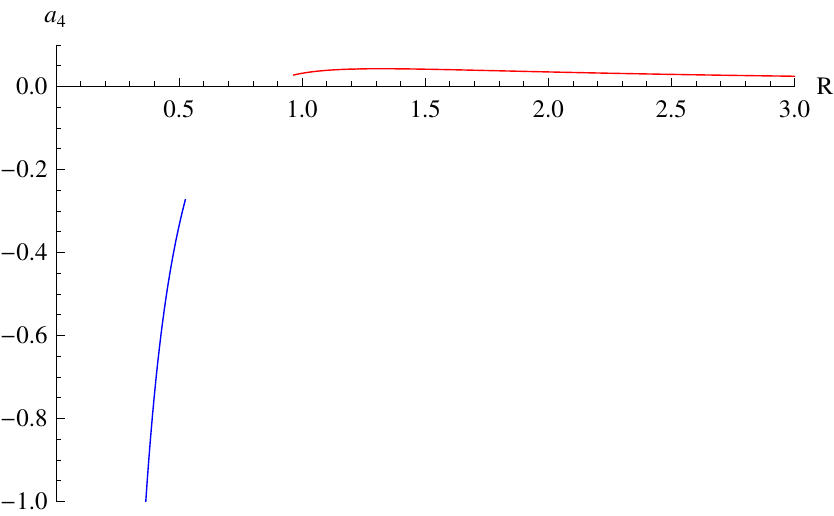}
\caption{ The $a_4(R)$ of  disk (blue)topology solutions for $0\le R\le
R_d=0.528$ and cylinder (red) topology solutions for
$R_{cyl}=0.968\le R$ with $\lambda_{GB}=0.05$, $z_0=1$. For $R_d<R<R_{cyl}$, the
solutions have cusp shape but are not unique. It suggests the absence of saddle point. We thus leave it open in the plot. } \label{hia4}
\end{figure}
%%%%%%%%%%%%%%%%%%%%%%%%%%%%%%%%%%%%%%%%%%

Plugging (\ref{hir}) into (\ref{hids}),  we obtain
\begin{equation}\label{dadrGBS}
\frac{d A}{dR}=-{\cal
H}(z_m)\frac{dz_m}{dR}+KRa_4(R)+\text{UV-dependent terms}+{\cal
O}(\epsilon),
\end{equation}
where
\begin{equation}
K=-L^3\frac{8\sqrt{2}\,(1-4\lambda_{GB})}{\sqrt{1+\gamma}\,(1+\gamma-2(2+\gamma)\lambda_{GB})}.
\label{k}
\end{equation}
and ${\cal O}(\epsilon)$ terms vanish at $\epsilon\to 0$ limit and
are not relevant. The UV-dependent terms are
\begin{equation}
\frac{c_1}{\epsilon^2}+\frac{c_2}{R^2}\log({
\epsilon \over R})+\frac{3c_2}{4R^2}\,
\end{equation}
where
\begin{equation}
c_1=L^3\frac{1+\gamma+4\lambda_{GB}}{\sqrt{2}\,(1+\gamma)^{3/2}},\qquad\qquad c_2 = -L^3\frac{\sqrt{1+\gamma}\,(1+\gamma-4\lambda_{GB})}{16\sqrt{2}}\,.
\end{equation}

%%%%%%%%%%%%%%%%%%%%%%%%%%%%%%%%%%%%%%%%
\begin{figure}[htbp]%[H]
\includegraphics[scale=1.00]{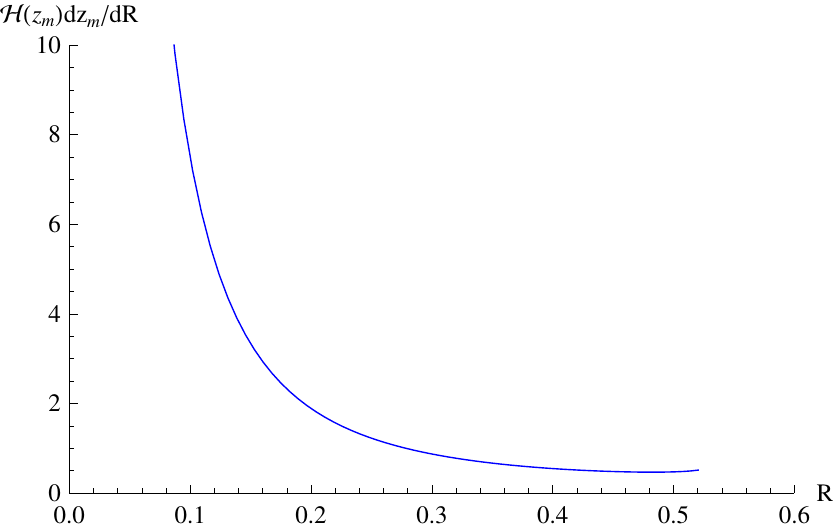}
\caption{${\cal H}(z_m)\frac{dz_m}{dR}$  with $\lambda_{GB}=0.05$, $z_0=1$
for disk topology solutions} \label{hd}
\end{figure}
%%%%%%%%%%%%%%%%%%%%%%%%%%%%%%%%%%%%%%%%%%

\begin{figure}[htbp]%[H]
\vspace{.3cm}
\includegraphics[scale=1.00]{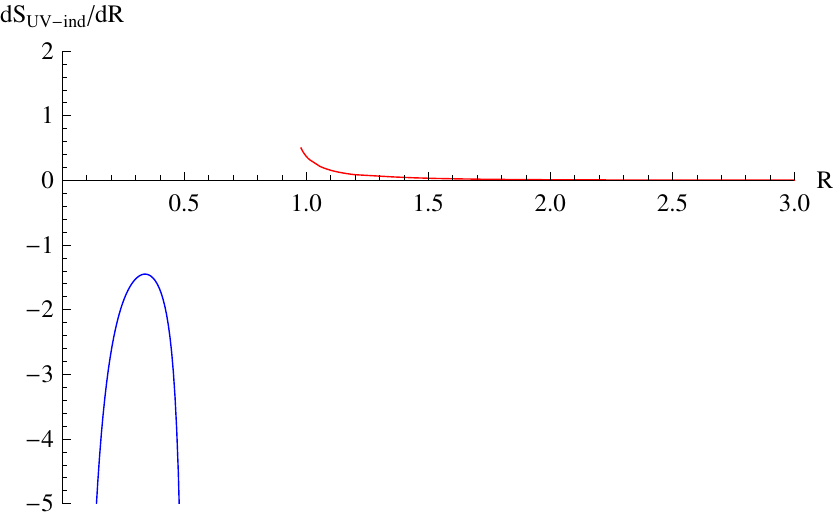}~~~~~
\caption{The ${dS_{\rm UV-ind}\over dR}$ for the disk (blue) topology
 for $0\le R\le R_d=0.528$ and the cylinder (red) topology
 for $R_{\rm cyl}=0.968\le R$ with $\lambda_{GB}=0.05$, $z_0=1$. For
$R_d<R<R_{\rm cyl}$, solutions have cusp shape. In this region, there
is no way to find the unique ${dS_{\rm UV-ind}\over dR}$ with fixed $R$.
}
\label{hidss}
\end{figure}

   We solve $a_4(R)$ and ${\cal H}(z_m)\frac{dz_m}{dR}$ (for disk topology only) numerically, and the results are given in Fig. \ref{hia4} and Fig. \ref{hd}, respectively. On the other hand,  for the cylinder topology, ${\cal H}(z_m)\frac{dz_m}{dR}=0$.  Using the above numerical data, we can then apply the same differential subtraction scheme given in \eq{RGAdS5} to extract from \eq{dadrGBS} the RG flow of the renormalized entanglement entropy in this case, and the numerical result is shown in Fig. \ref{hidss}.

%%%%%%%%%%%%%%%%
%%%%%%%%%%%%%%%%%%%%%%%%%%%%%%%%%%%%%%%%%%

Compared with Fig. \ref{figRGAdS5soliton} for the $\lambda_{GB}=0$ case, we find that the Gauss-Bonnet interaction
brings ambiguity to the transition between UV and IR regimes, since there appears a blank window between these two phases in which only non-unique cusp solutions exist. Despite this, the C-theorem still holds in the UV regime, and the feature that the renormalized entanglement entropy tends to constant in the IR regime is still retained.

\subsubsection{Extracting the topological entanglement entropy}

%%%%%%%%%%%%%%%%%%%%%%%%%%%%%%%%%%
In order to extract the topological entanglement entropy, we perform the large $R$ expansion
for the action (\ref{actionGB}) and equation of motion (\ref{eomGB}) as following:
\begin{multline} \label{actionepGB}
A = \int^{z_m}_\epsilon dz
\left\{ \frac{L^4 ( L_{\rm AdS}^2 (1 + 2 \lambda_{GB} f - \lambda_{GB} z {\dot f}) + L^2 f \,{\dot r}_1^2 ) }
{ L_{\rm AdS}^2 \, z^3 \sqrt{L_{\rm AdS}^2 + L^2 f \,{\dot r}_1^2} } R \right. \\
+ \frac{L^4}{ L_{\rm AdS}^2\,z^3\, (L_{\rm AdS}^2 + L^2 f {\dot r}_1^2)^{3/2} }
\left[ r_1 (L_{\rm AdS}^2 + L^2 f {\dot r}_1^2) (L_{\rm AdS}^2 (1 + 2 \lambda_{GB} f - \lambda_{GB} z {\dot f})
 + L^2 f {\dot r}_1^2 ) \right. \\
 + {\dot r}_1 \left( L_{\rm AdS}^4 \,\lambda_{GB}\, z^2 {\dot f} + L^2 f^2 ( L^2 {\dot r}_1^2 {\dot r_2}
- 2 L_{\rm AdS}^2 \,\lambda_{GB}\, (z {\dot r}_1^2 + {\dot r}_2)  ) \right. \\
\left. \left. \left. +\, L_{\rm AdS}^2\,f\, ( -2 L_{\rm AdS}^2 \,\lambda_{GB}\, z + L^2 ( {\dot r}_2
+ \lambda_{GB} z {\dot f} ( z {\dot r}_1^2 + {\dot r}_2) ) )\right) \right] + {\cal O}(\frac{1}{R})\right\}\,,
\end{multline}
\begin{multline}\label{eomexpandGB}
0 \,=\, \left\{ L^6 f^2 {\dot r}_1^5 (-6 f + z {\dot f})
+ L^4  L^2_{\rm AdS}\, f {\dot r}_1^2
\left(
12 \,\lambda_{GB} f^2 {\dot r}_1 + z \, {\dot f} \,{\dot r}_1 (\,3 - \lambda_{GB}\, z\, {\dot f})\right.\right.\\
\left.  + 8 \,\lambda_{GB}\, z\, f^2 {\ddot r }_1
 + 2\, f ( -\,6 \,{\dot r}_1 - 3 \,\lambda_{GB}\, z \, {\dot f} {\dot r}_1
 +\, z\, {\ddot r }_1 - 2 \,\lambda_{GB}\, z^2  {\dot f} {\ddot r }_1 + \lambda_{GB} \,z^2\, {\ddot f} {\dot r}_1 ) \right) \\
+ 2\, L^2 L^4_{\rm AdS} \left(\, z  {\dot f} {\dot r}_1
 + \lambda_{GB} \, z^2 {\dot f}^2 {\dot r}_1 + 2 \,\lambda_{GB}\, f^2 (\,3 \,{\dot r}_1 - z\, {\ddot r }_1\,)
 + f \,( -\,3\, {\dot r}_1 - 6 \,\lambda_{GB} \,z \, {\dot f} {\dot r}_1 \right. \\
\left.\left. +\, z \,{\ddot r}_1  + \lambda_{GB}\,z^2 {\dot f} {\ddot r}_1 + \lambda_{GB} \,z^2  {\ddot f}{\dot r}_1\, )
\right)\right\} R + {\cal O}( R^0)\,.
\end{multline}
For the cylinder topology which dominates at large $R$, we have ${\dot r}_1$ finite when $z \to 0$\,, for which equation (\ref{eomexpandGB}) gives
\be\label{r1eqz0GB}
{\dot r}_1(0) \left\{ \left[\, f_0^2\, {\dot r}_1(0)^2 +  ( 1 - \lambda_{GB} f_0 )\, \right]^2
-  \,\lambda_{GB}^2 f_0^2  \right\}= 0 \,.
\ee
Since the term in the curly braces of (\ref{r1eqz0GB}) are positive definite, we have ${\dot r}_1(0) = 0$. Note that $r_1(0) = 0$ and the fact that ${\dot r}_1 = 0$ is a solution of equation (\ref{eomexpandGB}). We then conclude that the unique solution to (\ref{eomexpandGB}) is $\,r_1 = 0\,$. Then, it is straightforward to see that the $R$-independent terms in (\ref{actionepGB}) vanish. This yields zero topological entanglement entropy. This is consistent with the expectation in \cite{Hayden:2011ag} that the topological order will not show up in the leading order of $1/N$ expansion, which captures up only classical phenomena and not the quantum ones such as the topological order.

\vspace*{4mm}

\subsection{Renormalized entanglement entropy for the Gauss-Bonnet corrected black hole}

  Now we turn to case of AdS$_5$ black hole with Gauss-Bonnet correction. The bulk theory is the same as for the Gauss-Bonnet corrected AdS$_5$ soliton, and the black hole metric is the doubled Wick rotation of \eq{metricGB}, which takes the form as
\begin{equation}
ds^2=\frac{L^2}{z^2}\left(-f(z)dt^2+\frac{1}{f(z)}dz^2+f_0(dr^2+r^2(d\theta^2+\sin^2\theta d\phi^2)\right)
\label{indads5bh}
\end{equation}
where, $f(z)$ and $f_0$ are the same as (\ref{f0def}).

By considering a disk on the boundary with radius $R$, the induced
metric of the minimal surface becomes
\begin{equation}\label{indads5bh2}
ds_{\rm ind}^2=L^2\left(\frac{1}{z^2}\left(\frac{1}{f(z)}+f_0\dot{r}(z)^2\right)dz^2+\frac{f_0r(z)^2}{z^2}(d\theta^2+ \sin^2\theta d\phi^2)\right)
\end{equation}
where $r$, $\theta$ and $\phi$ are radial, polar and azimuth
coordinates respectively.

The functional for the entanglement entropy is the same as \eq{heeGB}.  Using (\ref{indads5bh2})
we obtain \be \sqrt{h}\,(1+\lambda_{GB} L^2 {\cal R})=
\frac{2f_0r^2\sqrt{1+f_0f\dot{r}^2}}{z^3\sqrt{f}}+\lambda_{GB}\frac{4\left(z^2+f_0f(r^2-2zr\dot{r}+2z^2\dot{r}^2)\right)}{z^3\sqrt{f}\sqrt{1+f_0f\dot{r}^2}}+\dot{q},
\ee
where
\be
q(z)= \lambda_{GB} \frac{ 8  f_0 \sqrt{f} \,r (r - z \dot{r}) }{ z^2 \sqrt{ 1 + f_0 f \dot{r}^2} } \,.
\ee
Integrating the term $\dot{q} (z)$ in the bulk yields a surface term cancelling the Gibbons-Hawking term, then (\ref{heeGB}) becomes
\bea
A &=&\int_{\epsilon}^{z_m}dz\left(
\frac{2f_0r^2\sqrt{1+f_0f\dot{r}^2}}{z^3\sqrt{f}}+\lambda_{GB}\frac{4\left(z^2+f_0f(r^2-2zr\dot{r}+2z^2\dot{r}^2)\right)}{z^3\sqrt{f}\sqrt{1+f_0f\dot{r}^2}}\right), \nn \\
&:=& \int dz \; \mathcal{L},
\label{adbh}
\eea
from which we derive the equation of motion for $r(z)$ as follows
\begin{multline}
0=\frac{1}{z^4\sqrt{f}(1+f_0f\dot{r}^2)^{5/2}}f_0(-6z^2\lambda_{GB}(z\dot{f}\dot{r}-2f_0f^2\dot{r}^3-2f(\dot{r}-z\ddot{r}))
+4zr(1+z\lambda_{GB}\dot{f}\\
+4\lambda_{GB}f_0^2f^3\dot{r}^4-2f(\lambda_{GB}+f_0(-1-z\lambda_{GB}\dot{f})\dot{r}^2)
+f_0f^2\dot{r}(2\lambda_{GB}\dot{r}+f_0\dot{r}^3-6z\lambda_{GB}\ddot{r}))\\
+f_0r^2(-z\dot{f}\dot{r}+f(6(1+z\lambda_{GB}\dot{f})\dot{r}-zf_0\dot{f}\dot{r}^3-2z\ddot{r})
+2f_0f^3\dot{r}^2(-6\lambda_{GB}\dot{r}+3f_0\dot{r}^3-4z\lambda_{GB}\ddot{r})\\
-2f^2(6\lambda_{GB}\dot{r}-6f_0\dot{r}^3-2z\lambda_{GB}\ddot{r}+zf_0\dot{f}^2\ddot{r}))).
\label{ads5bheom}
\end{multline}
The UV behavior of the solution $r(z)$ is obtained as
\begin{equation}
r(z)=R+\frac{\lambda_{GB}(-1-\gamma+4\lambda_{GB})}{2R(-1+\gamma+4\lambda_{GB})}z^2+a_4(R)z^4+\cdots
\label{adsbhuvr}
\end{equation}
where $\gamma$ is defined in (\ref{higm}) and $a_4(R)$ should be determined by solving
the full equation of motion (\ref{ads5bheom}).

  Unlike the complication for the Gauss-Bonnet AdS$_5$ soliton case, there are well-defined solutions of disk topology for all $R$. As for the AdS$_4$ black hole, we just need to consider the disk topology. The RG flow of the on-shell action is given by (\ref{hids}), and we need to see if the first term in \eq{hids} has no zero contribution or not.  From \eq{adbh} we can obtain
\begin{equation}
\Pi:=\frac{\delta
\mathcal{L}}{\delta\dot{r}}=\frac{2f_0\sqrt{f}(-4z\lambda_{GB}r+2z^2\lambda_{GB}\dot{r}(3+2f_0f\dot{r}^2)+f_0r^2\dot{r}^2(1+f(-2\lambda_{GB}+f_0\dot{r}^2)))}{z^3(1+f_0f\dot{r}^2)^{3/2}}
\label{adsbhpi}
\end{equation}
and
\begin{equation}
\mathcal{H}:=\Pi
\dot{r}-\mathcal{L}=-\frac{2(2z^2\lambda_{GB}-4z\lambda_{GB}f_0^2f^2r\dot{r}^3+f_0r^2(1+4\lambda_{GB}f_0f^2\dot{r}^2+f(2\lambda_{GB}+f_0\dot{r}^2)))}{z^3\sqrt{f}(1+f_0f\dot{r}^2)^{3/2}}
.\label{adsbhham}
\end{equation}
For disk topology,
\be \label{adsrzm1}
\frac{dr}{dz}|_{z=z_m}=\infty, \qquad
r(z_m)=0.
\ee
By plugging (\ref{adsrzm1}) into (\ref{adsbhham}), we get
\begin{equation}
\mathcal{H}(z_m)=\frac{8\lambda_{GB}\sqrt{f_0}r(z_m)}{z_m^2}=0.
\end{equation}
Then the first term of (\ref{hids}) becomes zero.
 
From \eq{hids},  (\ref{adsbhuvr}) and (\ref{adsbhpi}), we get
\begin{eqnarray}\label{RGhighBH}
\frac{d A}{dR}&=&\frac{1}{R\sqrt{2}\gamma\lambda_{GB}^2}\sqrt{\frac{1-\gamma}{\lambda_{GB}}}[-\lambda_{GB}^2\left(1+\gamma-2(4+3\gamma)\lambda_{GB}+16\lambda_{GB}^2\right)\nonumber\\
&&-4R^3(-1+\gamma+2\lambda_{GB})(-1+4\lambda_{GB})a_4(R)]\nonumber\\
 &+& \text{UV-dependent terms}+{\cal O}(\epsilon)
\end{eqnarray}
where the UV-dependent divergent terms becomes
\begin{equation}
R\sqrt{\frac{2-2\gamma}{\lambda_{GB}}}\frac{(1+\gamma+4(-1+\gamma)\lambda_{GB})}{(1+\gamma-4\lambda_{GB})}\frac{1}{\epsilon^2}.
\end{equation}
In fact in the action there is an additional R-independent logarithmic UV divergent term, which does not appear in (\ref{RGhighBH}).
To see this, we substitute the UV expansion of $r(z)$ (\ref{adsbhuvr}) into the action (\ref{adbh}) and find it as
\be
\sqrt{\frac{\lambda_{GB}}{ 2- 2 \gamma}}\,\, ( 1 + \gamma - 12 \lambda_{GB}) \log{\frac{\epsilon}{R}}\,,
\ee
which should be subtracted along with the quadratic divergence when evaluating $S_{\rm finite}$.

\begin{figure}[b]
\begin{center}
\includegraphics[scale=1.00]{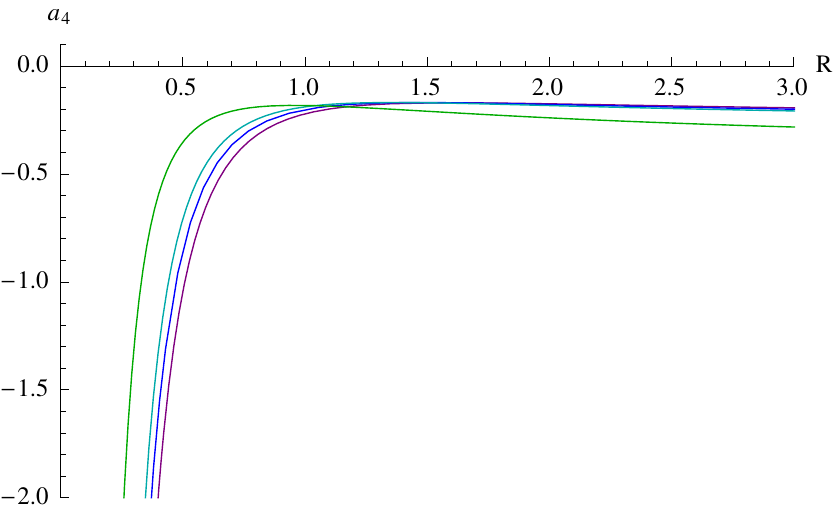}
\end{center}
\caption{ The $a_4(R)$'s for Gauss-Bonnet corrected AdS$_5$ black holes with $z_0 = 1$ and $\lambda_{GB} =$ 0 (purple),
0.05 (blue), 0.09 (cyan), 0.2 (green), respectively. } \label{figa4high}
\end{figure}

\begin{figure}[t]
\begin{center}
\includegraphics[scale=1.00]{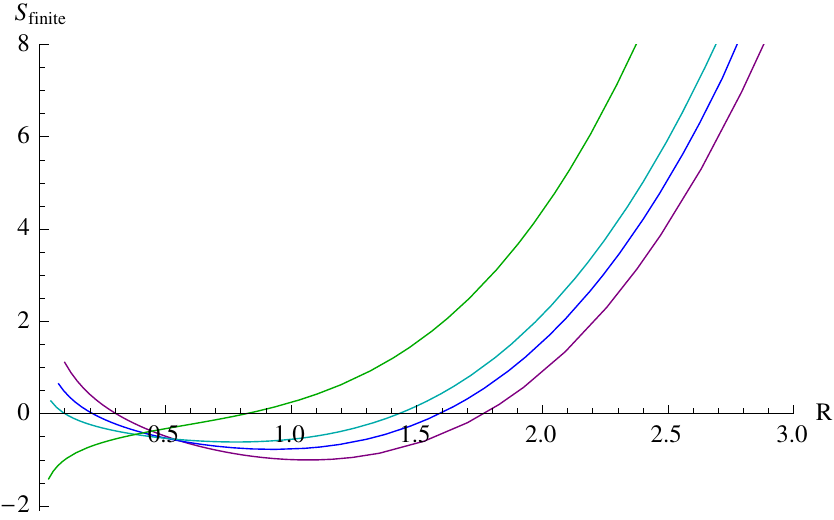}
\end{center}
\caption{The $S_{\rm finite}$'s for Gauss-Bonnet corrected AdS$_5$ black holes with $z_0 = 1$ and $\lambda_{GB} =$ 0 (purple),
0.05 (blue), 0.09 (cyan), 0.2 (green), respectively. } \label{figShigh}
\end{figure}

\begin{figure}[t]
\begin{center}
\includegraphics[scale=0.8]{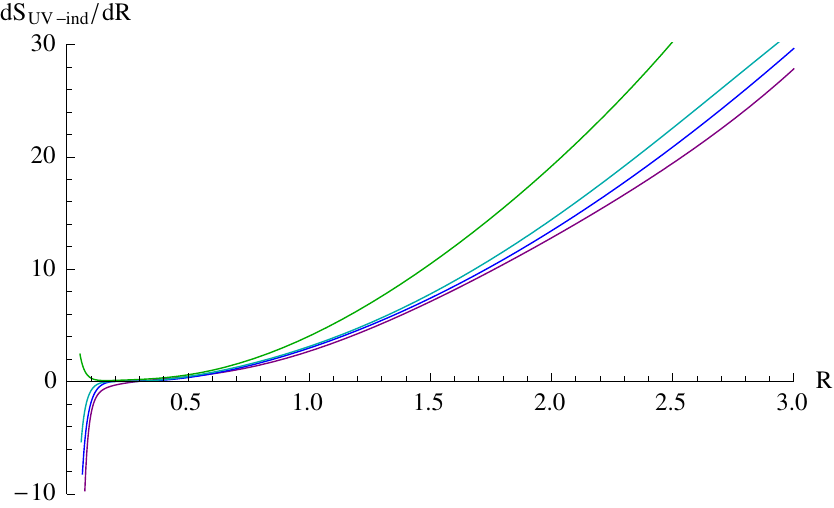}~~~~~
\includegraphics[scale=0.8]{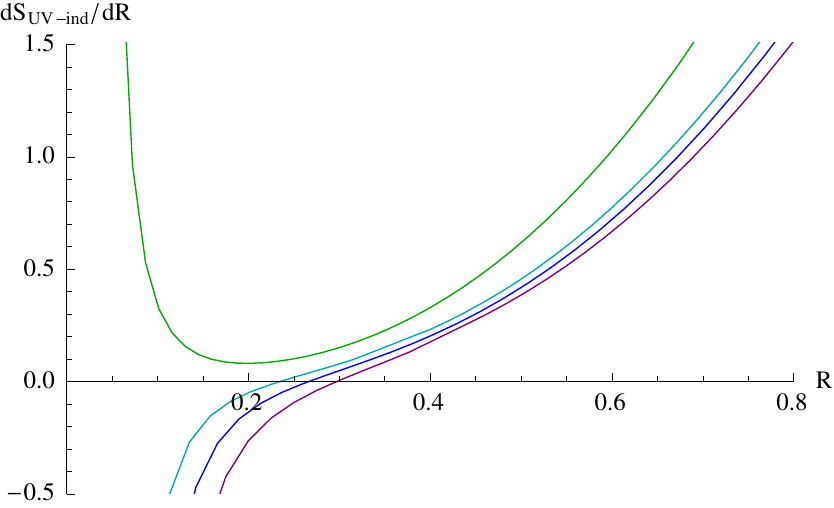}
\end{center}
\caption{Left: The $dS_{\rm UV-ind} \over dR$'s for Gauss-Bonnet corrected AdS$_5$ black holes with $z_0 = 1$ and $\lambda_{GB} =$ 0 (purple),
0.05 (blue), 0.09 (cyan), 0.2 (green), respectively. Right: Zoom-in of the region with the sign-change.} \label{figRGhigh}
\end{figure}

We then numerically solve $a_4(R)$ and the finite part of the on-shell action $S_{\rm finite}$ for different values of $0 \le \lambda_{GB} \le 1/4$, and the results are plotted in Fig.\,\ref{figa4high} and Fig.\,\ref{figShigh}, respectively. To extract the RG flow $\frac{dS_{\rm UV-ind}}{dR}$ from \eq{RGhighBH}, we again apply the differential subtraction scheme based on \cite{Liu:2012ee}. Explicitly, it is
\be
\frac{dS_{\rm UV-ind}}{dR}= {1\over 2} (R\partial_R+1)(R\partial_R-1) {d A\over dR}.
\ee
The numerical results  are  shown in Fig.\,\ref{figRGhigh}.

  The Gauss-Bonnet interaction corresponds to some operator at the sub-leading order in the inverse 't Hooft coupling expansion. It denotes the finite coupling correction to the infinite 't Hooft coupling limit in the dual field theory.  First, we notice that for $\lambda_{GB}=0$, i.e., corresponds to the AdS$_5$ black hole of Einstein gravity, the UV behavior of $\frac{dS_{\rm UV-ind}}{dR}$ is different from the one for the AdS$_4$ black hole case. In the latter case, the $\frac{dS_{\rm UV-ind}}{dR}$ is always positive, but here it is negative at UV and then turn to become positive to capture the volume law nature at IR.    We can see that the effect of the the Gauss-Bonnet interaction is to change the UV behavior of $\frac{dS_{\rm UV-ind}}{dR}$ so that it becomes all positive even at UV for large enough $\lambda_{GB}$.

  Despite the discrepancies in the UV behaviors for different $\lambda_{GB}$'s, the qualitative behaviors of the results are the same: the on-shell actions catch the volume law of the thermal entropy in a smooth way as $R$ becomes large, just like the AdS$_4$ black hole case. This again supports the postulate proposed in \cite{Swingle:2011mk}. We thus conclude that the crossover is not effected by turning on the Gauss-Bonnet interaction.

In \cite{Brigante:2008gz,Buchel:2009tt} it is pointed out that the holographic dual field theory with $\lambda_{GB} > {9 \over 100}$ will violate microscopic causality, however, although the small $R$ behaviors of $S_{\rm finite}$ and RG flow become quite different for sufficiently large $\lambda_{GB}$, e.g., $\lambda_{GB}=0.2$ in Fig.\,\ref{figShigh} and Fig.\,\ref{figRGhigh}, nothing exotic happens in this regime. This agrees with the same consideration for the Gauss-Bonnet corrected AdS$_5$ soliton in \cite{Ogawa:2011fw}.  However, there are some concern about the relation between the quantum entanglement and the causality formulated from the consideration of the quantum information sciences \cite{IC,IC1}, it may deserve further study to understand this issue in the context of holographic entanglement entropy.

\begin{figure}[h]
\begin{center}
\includegraphics[scale=0.80]{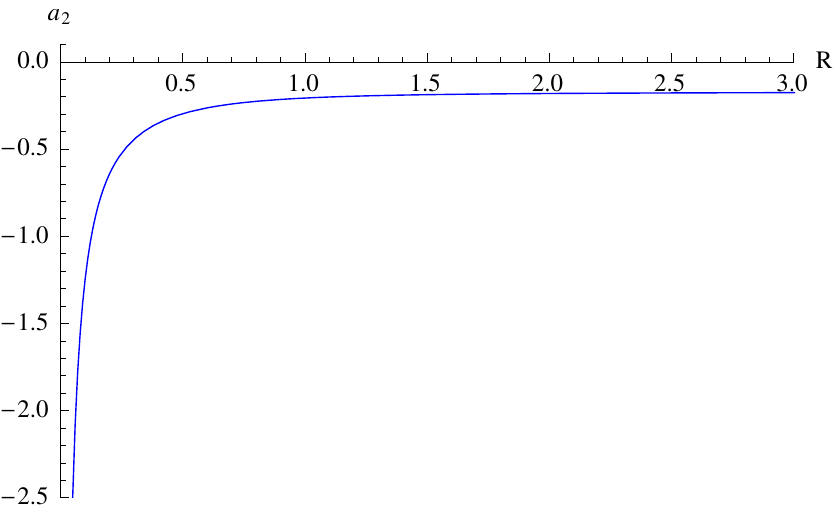}~~~~~
\includegraphics[scale=0.80]{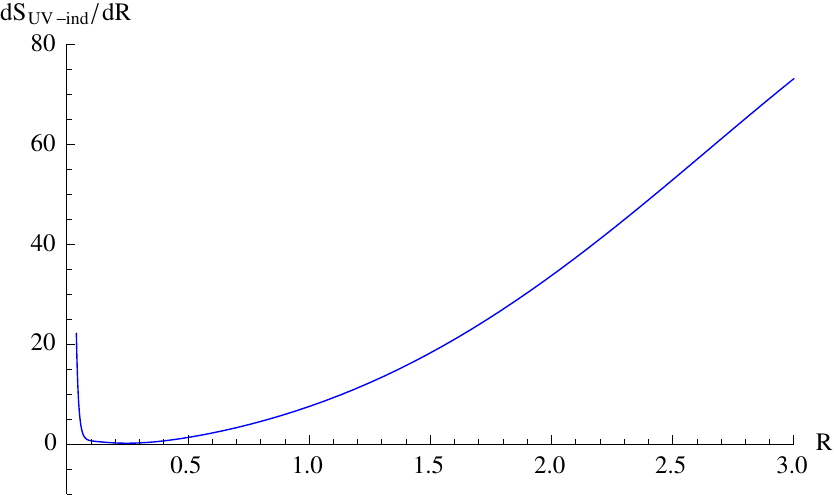}
\end{center}
\caption{Left: The $a_2(R)$ for $\lambda_{GB}=1/4$ and $z_0=1$. Right: The corresponding RG flow of the refinement.}
\label{a2025}
\end{figure}

Finally, we would like to give the numerical results for the $\lambda_{GB}=1/4$ case, for which the viscosity to entropy ratio vanishes for the holographic dual field theory.  The peculiar feature of the geometry is the harmonic function $f(z)$ becomes $1-({z\over z_0})^2$, which is quite different from the one for asymptotic AdS$_5$, namely, $1-({z\over z_0})^4$. The UV behavior of the solution is different from the $\lambda_{GB}<1/4$ cases and is given by
\be\label{rz025}
r(z)=R+a_2 z^2+ {a_2 z^4 (3(z_0^2+16a_2^2 z_0^4)+64R (a_2z_0^2 +12 a_2^3 z_0^4)+4 R^2 (1+96 a_2^2 z_0^2 +768 a_2^4 z_0^4))\over 4z_0^2 (3z_0^2 +48 a_2 R z_0^2 +4 R^2 (1+48 a_2^2 z_0^2))} + \cdot\cdot\cdot
\ee
where $a_2(R)$ instead of $a_4(R)$ should be determined by solving the full equation of motion, and the result is shown in the left plot of Fig. \ref{a2025}. From  \eq{rz025} and \eq{hids} we can obtain
\be
{d A\over dR}={4\sqrt{2} R \over \epsilon^2} -2\left(\sqrt{2}((4+{8R^2 \over z_0^2}) a_2 +48R a_2^2 +128 R^2 a_2^3 + R ({1\over z_0^2}-2 {d a_2 \over dR}))\right) + \mathcal{O}(\epsilon^2).
\ee
The result of the RG flow is shown in the right plot of Fig. \ref{a2025}, from which we see again the volume law for large $R$ as well as the crossover.

\section{Conclusions and Discussions: IR fixed-point state from AdS/MERA}\label{MERAsec}

   In this paper, we have considered the refinement of the holographic entanglement entropy and its RG flow behavior for the systems dual to AdS solitons and black holes. The holographic entanglement entropy for AdS solitons has different scaling behavior from AdS cases, so does the UV-independent piece, i.e., the renormalized entanglement entropy.     As for the cases of AdS black holes, our results yield the refined version of the conjecture given in \cite{Swingle:2011mk} that the transition of the UV cutoff-independent piece of the entanglement entropy between the IR and UV regimes is a smooth crossover even with the correction of the Gauss-Bonnet interaction.  On the other hand, for the AdS soliton cases,  we find that the renormalized entanglement entropy for $d=4,5$ is not monotonically decreasing along the RG flow, nor are they always positive definite. In $d=4$ case, such behavior is related to the geometry of the entangling surface (torus) which singles out the B-type anomaly and there is no conflict with the C-theorem. Generally, one should expect the renormalized entanglement entropy to play the role of a C-function when the entangling surface is spherical only \cite{Liu:2012ee}.
  
      Turning on the Gauss-Bonnet interaction will make the region around the confinement phase transition become ill-defined. Similarly, the irrelevance of the Gauss-Bonnet interaction to the topological entanglement entropy is also checked so that there is no non-trivial topological order for the AdS$_5$ soliton and its Gauss-Bonnet cousins.

   Before ending this paper, we would like to devote the rest of the discussions on how to understand the entangling nature of the IR fixed-point states of the holographic dual theory based on the conjecture of AdS/MERA proposed in \cite{Swingle:2009bg,EvenblyVidal}. We will argue that non-extremal AdS soliton has the product state as its IR fixed-point state, and the extremal AdS soliton instead has the nontrivial entangled state as the IR fixed-point state. The different nature of the IR fixed-point states depends on the topology of the large R entangling hypersurfaces. If our arguments here hold, this may be seen as another triumph of AdS/CFT in using the simple geometric picture to characterize the entangled mean field states. Further development along this line may reveal the holographic and geometric classification of the topologically ordered phases in the strongly interacting condensed matter systems.

    Though the wave function of a many-body system could look quite complicated, it could be simplified a lot through some appropriate local unitary operations, especially when these operations are adopted to remove short-range entanglement among neighboring particles. An example of such unitary operations is the $CZ$ (controlled-Z) operation, which transforms a Bell state into product state as
\be
CZ(|0\rangle|+\rangle+|1\rangle|-\rangle)=(|0\rangle+|1\rangle)|+\rangle=\sqrt{2}|+\rangle|+\rangle,
\ee
where $|\pm\rangle={1\over \sqrt{2}}(|0\rangle \pm |1\rangle)$. Moreover, if we are only interested in the low energy behaviors of the system, we could further coarse-grain the wave function by merging the the neighboring sites after removing the short-range entanglement. After repeating the above two steps, we will obtain a far more simple wave function at the IR fixed-point, or the so-called mean field state. This is the so-called quantum state RG transformation \cite{qsrg1,qsrg2} (see also \cite{qsrg3,qsrg4} for practical numerical study) as shown in Fig. \ref{Fig1}, and can be adopted to classify the phases of the many-body systems.  That is, all the wave functions flowing to the same fixed-point state under quantum state RG transformation describe the same phase. According to this scheme of classification, for gapped systems one may expect two kinds of the IR fixed-point states. One is the product state which encodes no quantum entanglement. The other kind is the nontrivial topological ordered states, which encode either long-range entanglement or some short-range entanglement protected by symmetries \cite{qsrg2}.  In this way, one can tell which phase the system belongs to by looking into the IR fixed-point wave function, instead of the UV ones. In other words, the gapped systems are classified by the patterns of the quantum entanglement of the IR fixed-point states. Especially, for 1-dimensional spin chain, it was shown that all the ground states will flow to trivial product state under generic quantum state RG transformation unless some symmetries are preserved during the RG flow \cite{SRE1,SRE2,SRE3,SRE4,SRE5}. However, the classification of higher dimensional systems are still under development. The above scheme of looking into the IR fixed-point state is in contrast to what has been adopted in this paper and summarized in \eq{SUV} by looking into the UV scaling behaviors of the entanglement entropy for the relativistic CFTs.

%%%%%%%%%%%%%%%%%%%%%%%%%%%%%%%%%%%%%%%%
\begin{figure}[htbp]%[H]
%\vspace{.3cm}
\includegraphics[width=17cm,height=12cm]{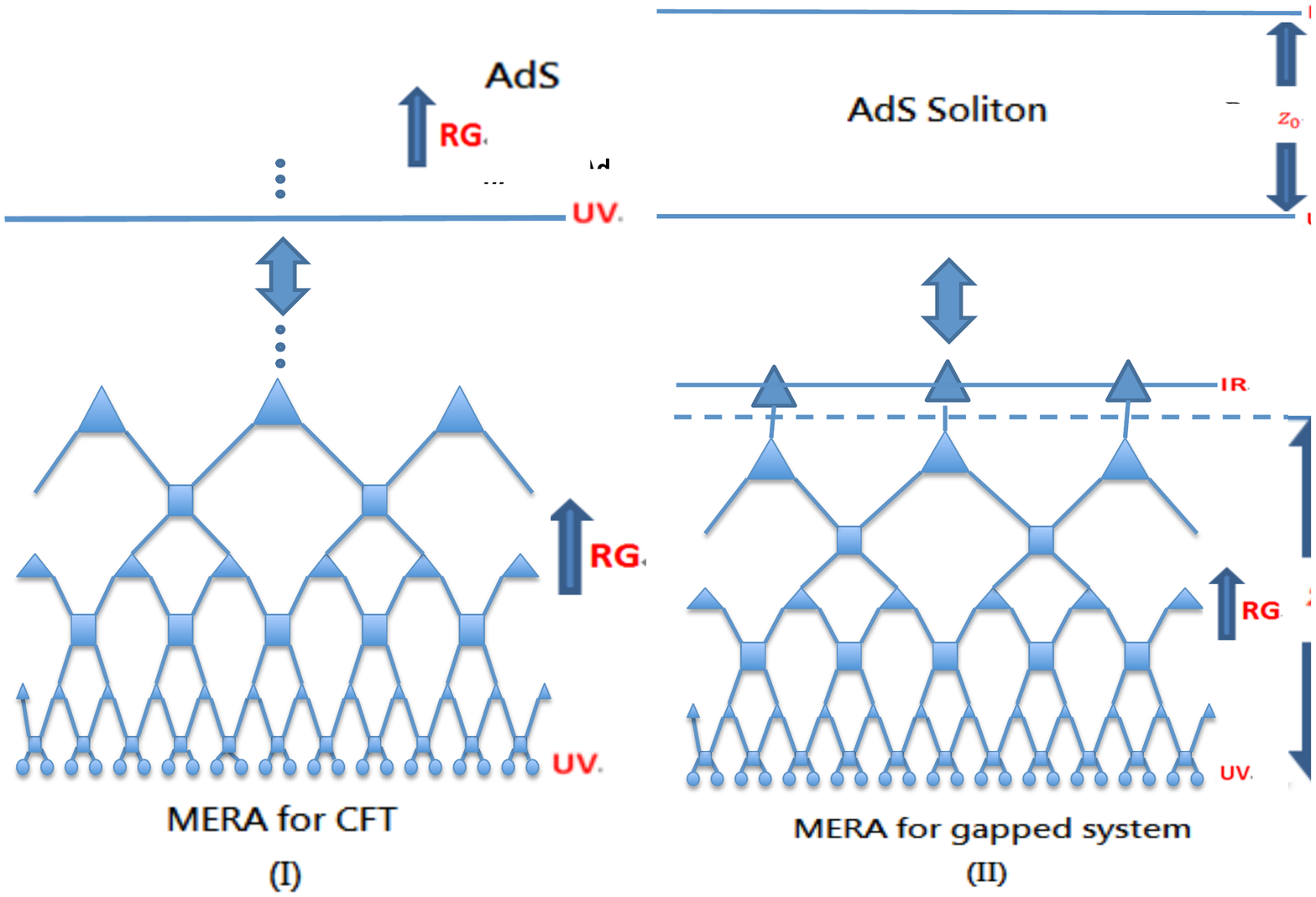}~~~~~
\caption{MERA network and its dual AdS geometry.  Here the disentanglers are denoted by solid squares, and the isometries by the solid triangles. The links at different levels encode short-range entanglement at different scales. (I) The MERA for CFT and its dual is the AdS space. Note that the depth of the MERA can be extended indefinitely as its dual AdS geometry. (II) MERA for gapped system and it dual is the AdS soliton. The MERA and its dual geometry end at some IR scale.  For simplicity, we just plot the  one-dimensional MERA, however, it is straightforward to plot for higher dimensional cases.} \label{mera1}
\end{figure}
%%%%%%%%%%%%%%%%%%%%%%%%%%%%%%%%%%%%%%%%%%

The local unitary operation and the coarse-graining in the quantum state RG transformation can be implemented as the quantum gates of the quantum circuit with some pre-prepared inputs. Therefore, the whole procedure can be viewed as some time evolving procedure and then be implemented to solve some many-body systems.   This idea then results in algorithm of multi-scale entanglement renormalization ansatz (MERA) \cite{MERA}, and see \cite{Evenbly} for more detailed introduction. In MERA, the local unitary operations in removing the short-range entanglement are called disentanglers, and the merging operations for coarse-graining are called isometries. Then, the whole procedure of quantum state RG transformation can be piled up as a network of disentanglers or isometries. The depth of the MERA network can be thought as the time evolution or RG flow, and the links in the network denote the short-range entanglement among the neighboring sites. A typical MERA network for both CFT and gapped system are depicted in Fig. \ref{mera1}.  Note that the depth for the CFT is indefinite due to the scaling invariance and could be infinite for an infinite UV system. On the other hand, the depth for the gapped system is finite as the RG procedure must end when reaching the IR mass gap.

   In practical, the MERA can be used to solve the ground state of the system by treating the disentanglers and isometries as the variational ansatz, which can then be determined by minimizing the expectation values of the Hamiltonian. For examples, see \cite{Evenbly} for this kind of applications.

   On the other hand, the MERA network yields a geometric picture of the quantum state RG, and indeed the geometry can be characterized by the aspect ratio of depth to width, i.e., $z\sim \log |\vec{x}|$. This aspect ratio encodes the block decimation of coarse-graining and is roughly coincident with the AdS geometry as first observed in \cite{Swingle:2009bg} and made more precise later in \cite{EvenblyVidal}.  For the gapped system, the finite depth is consistent with the geometry of AdS soliton with $z_0 \sim \log \xi$ where $\xi$ is the correlation length. Moreover, by utilizing the unitarity feature of disentanglers and isometries in the MERA network one finds that a site is only affected by the sites within its causal cone. The correlation between two distant sites are encoded by the intersection of the causal cones, which is pretty much the same as the geodesic in the AdS bulk connecting two boundary points. This then reminds the prescription of evaluating the boundary correlation functions in the AdS/CFT correspondence \cite{GKPW}. By the aspect ratio of depth to width, the length of the intersecting causal cone then yields the expected power law for CFT correlation function and the exponential decay behavior for the gapped one.

%%%%%%%%%%%%%%%%%%%%%%%%%%%%%%%%%%%%%%%%
\begin{figure}[htbp]%[H]
%\vspace{.3cm}
\includegraphics[width=17cm,height=14.5cm]{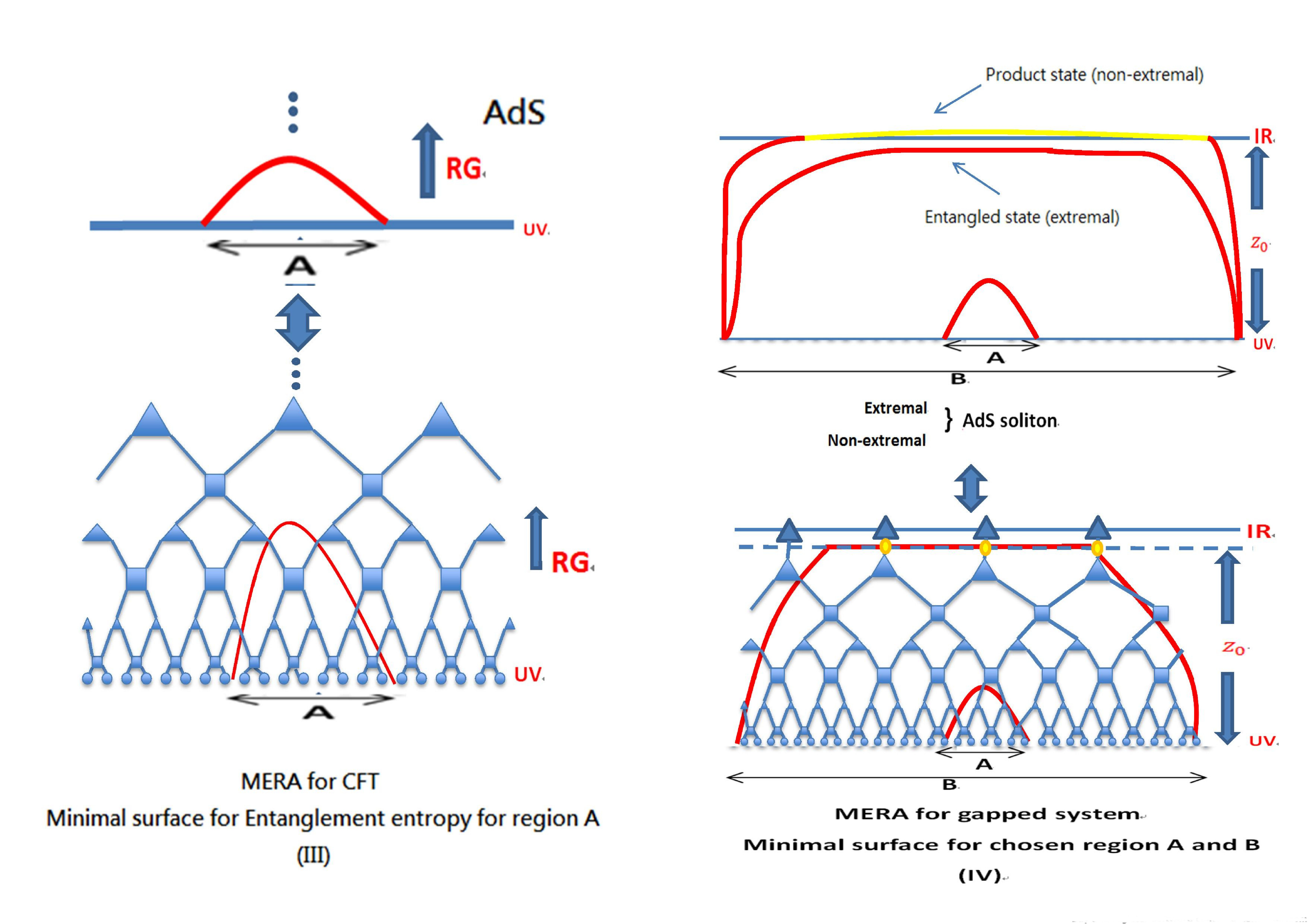}
\caption{Minimal surfaces for the entanglement entropy in the MERA and in its dual geometry. The entanglement entropy is obtained by counting the links which intersect the minimal surface. This implies that the entanglement entropy is contributed by the short-range entanglement at all length scales smaller than the linear size $R$ of the chosen region. (III) For the CFT case, the minimal surface is always in the disk topology.  (IV) For the gapped systems dual to non-extremal AdS soliton, the topology of the minimal surface changes from the disk at smaller $R$ to the cylinder at large $R$. Compare the minimal surfaces for MERA and AdS soliton, we conclude that the IR fixed-point state (the yellow part excluded from the minimal surface) is a product state since the links at the top level of MERA have no counterparts at the geometry side.
On the other hand, for the extremal AdS soliton, the minimal surface is always in disk topology, this is because the extremal AdS soliton has gapless KK modes which may retain the entangled pairs at the IR regime.
} \label{mera2}
\end{figure}
%%%%%%%%%%%%%%%%%%%%%%%%%%%%%%%%%%%%%%%%%%

    Similarly, the geometric picture of the holographic entanglement entropy is encoded in the minimal surface covering the boundary sites inside the chosen region as depicted in Fig. \ref{mera2} for both CFT and gapped systems (see also Fig. \ref{Fig1}). The entanglement entropy is proportional to the number of the links intersecting with the minimal surface because the links carry the short-range entanglement between the sites just inside and outside the chosen region. This then results in the expected area law for both CFT \footnote{It can also recover the logarithmic behavior for the $1+1$ CFT.} and gapped systems.  The most interesting point is that the link at different depth level of the MERA network actually encodes the short-range entanglement at the corresponding scale. To be more specific, the links at level 0 (the UV boundary) encode the short-range entanglement between nearest neighboring sites, but the links at level 1 encode the short-range entanglement between the next-nearest neighboring sites with the distance measured by the UV scale.   Therefore, MERA network geometrically and systematically displays how the short-range entanglements of different length scales are contributed to the total entanglement entropy of a chosen region at UV level.

        Especially, for the gapped system there exists a top layer in the MERA network, which represents the IR fixed-point and also encodes the short-range entanglement of the IR fixed-point state. Therefore, if the fixed-point state is not a product state, its short-range entanglement will contribute to the total entanglement entropy. Otherwise, there is nothing to contribute for a product state. This then corresponds to the following geometric picture.  Due to the existence of the IR top layer, the minimal surface covering the chosen region will have a flat bend-over near the top layer. If the fixed-point state is the product state, then the flat bend-over region of the minimal surface collect no entanglement from the fixed-point state. In this case, flat bend-over region can be effectively removed, and the resultant minimal surface can be effectively viewed as ending on the top-layer. This is indeed the IR dominating cylinder topology found in the non-extremal AdS soliton case.  From our above argument, it implies that the IR fixed-point state is the product state.  This result is consistent with the vanishing topological entanglement entropy\footnote{We restrict our discussions here for the $AdS_5$ soliton case, which is dual to the 2+1 gapped system.  On the other hand, the nature of the topological entanglement in higher dimensional system is not clear.} and the negative value of the finite part of the holographic entanglement entropy, which could compensate the positive UV contribution to make zero total entanglement entropy near IR fixed-point.

        On the other hand, for the extremal AdS soliton case we see that only disk topology exists so that the flat bend-over region does contribute to the holographic entanglement entropy. From the above argument, this could imply that the IR fixed-point state may not be the product state but a nontrivial entangled state.  Geometrically, the difference between extremal and non-extremal AdS soliton is that the spectator U(1) cycle for the former becomes non-compact at the IR fixed-point. That is, the IR fixed-point is a gapless state as the KK modes become massless at the extremal limit. These gapless excitations may retain some entangled pairs at the IR scale indicated by the flat bend-over region.

         The above speculation of the entangled properties of the IR fixed-point state from AdS/MERA can be further exemplified by our study of the AdS black hole. In this case, the AdS geometry provides more useful information than MERA, whose finite temperature version is barely studied.  Based on AdS/MERA, the finite temperature MERA network of the CFT is no longer extended indefinitely but will be terminated by the IR scale fixed by the temperature. This will be a helpful guideline when implementing the finite temperature MERA for CFT.  Moreover, from our numerical study we see that the the dominant topology at large $R$ is the disk one whose refined holographic entanglement entropy captures the volume law of the thermal entropy. According to the same consideration as for the AdS soliton case, this implies that the IR fixed-point state has nontrivial entanglement at IR scale. Indeed, the IR fixed point state should be a thermally mixed state and is different from the product state dual to the cylinder topology. Though we may need the pattern of thermal MERA to understand the how the multi-scale entanglements distribute at nonzero temperature.

\section*{Acknowledgements}
FLL thanks Xie Chen, Ching-Yu Huang, Wei Li, Hong Liu and Frank Pollmann for discussions, especially thanks Tadashi Takayanagi for classifying some subtle issues relevant to our revision.  We thank Pei-Hua Liu for drawing the MERA figures. FLL is supported by Taiwan's NSC grants (grant NO. 100-2811-M-003-011 and 100-2918-I-003-008). We thank the support of NCTS.


\begin{thebibliography}{99}

%\cite{Amico:2007ag}
\bibitem{Amico:2007ag}
  L.~Amico, R.~Fazio, A.~Osterloh and V.~Vedral,
  ``Entanglement in many-body systems,''
  Rev.\ Mod.\ Phys.\  {\bf 80}, 517 (2008)
  [quant-ph/0703044 [QUANT-PH]].
  %%CITATION = QUANT-PH/0703044;%%


%\cite{Calabrese:2009qy}
\bibitem{Calabrese:2009qy}
  P.~Calabrese and J.~Cardy,
  ``Entanglement entropy and conformal field theory,''
  J.\ Phys.\ A A {\bf 42}, 504005 (2009)
  [arXiv:0905.4013 [cond-mat.stat-mech]].
  %%CITATION = ARXIV:0905.4013;%%


  %\cite{Levin:2006zz}
\bibitem{Levin:2006zz}
  M.~Levin and X.~-G.~Wen,
  ``Detecting Topological Order in a Ground State Wave Function,''
  Phys.\ Rev.\ Lett.\  {\bf 96}, 110405 (2006).
  %%CITATION = PRLTA,96,110405;%%

 %\cite{Kitaev:2005dm}
\bibitem{Kitaev:2005dm}
  A.~Kitaev and J.~Preskill,
  ``Topological entanglement entropy,''
  Phys.\ Rev.\ Lett.\  {\bf 96}, 110404 (2006)
  [hep-th/0510092].
  %%CITATION = HEP-TH/0510092;%%


%\cite{Srednicki:1993im}
\bibitem{Srednicki:1993im}
  M.~Srednicki,
  ``Entropy and area,''
  Phys.\ Rev.\ Lett.\  {\bf 71}, 666 (1993)
  [hep-th/9303048].
  %%CITATION = HEP-TH/9303048;%%

   %\cite{Eisert:2008ur}
\bibitem{Eisert:2008ur}
  J.~Eisert, M.~Cramer and M.~B.~Plenio,
  ``Area laws for the entanglement entropy - a review,''
  Rev.\ Mod.\ Phys.\  {\bf 82}, 277 (2010)
  [arXiv:0808.3773 [quant-ph]].
  %%CITATION = ARXIV:0808.3773;%%


%\cite{Callan:1994py}
\bibitem{Callan:1994py}
  C.~G.~Callan, Jr. and F.~Wilczek,
  ``On geometric entropy,''
  Phys.\ Lett.\ B {\bf 333}, 55 (1994)
  [hep-th/9401072].
  %%CITATION = HEP-TH/9401072;%%

%\cite{Holzhey:1994we}
\bibitem{Holzhey:1994we}
  C.~Holzhey, F.~Larsen and F.~Wilczek,
  ``Geometric and renormalized entropy in conformal field theory,''
  Nucl.\ Phys.\ B {\bf 424}, 443 (1994)
  [hep-th/9403108].
  %%CITATION = HEP-TH/9403108;%%


%\cite{Ryu:2006bv}
\bibitem{Ryu:2006bv}
  S.~Ryu and T.~Takayanagi,
  ``Holographic derivation of entanglement entropy from AdS/CFT,''
  Phys.\ Rev.\ Lett.\  {\bf 96}, 181602 (2006)
  [hep-th/0603001].
  %%CITATION = HEP-TH/0603001;%%


%\cite{Ryu:2006ef}
\bibitem{Ryu:2006ef}
  S.~Ryu and T.~Takayanagi,
  ``Aspects of Holographic Entanglement Entropy,''
  JHEP {\bf 0608}, 045 (2006)
  [hep-th/0605073].
  %%CITATION = HEP-TH/0605073;%%


%\cite{Nishioka:2009un}
\bibitem{Nishioka:2009un}
  T.~Nishioka, S.~Ryu and T.~Takayanagi,
  ``Holographic Entanglement Entropy: An Overview,''
  J.\ Phys.\ A A {\bf 42}, 504008 (2009)
  [arXiv:0905.0932 [hep-th]].
  %%CITATION = ARXIV:0905.0932;%%

%\cite{Witten:1998zw}
\bibitem{Witten:1998zw}
  E.~Witten,
  ``Anti-de Sitter space, thermal phase transition, and confinement in gauge theories,''
  Adv.\ Theor.\ Math.\ Phys.\  {\bf 2}, 505 (1998)
  [hep-th/9803131].
  %%CITATION = HEP-TH/9803131;%%

%\cite{Nishioka:2006gr}
\bibitem{Nishioka:2006gr}
  T.~Nishioka and T.~Takayanagi,
  ``AdS Bubbles, Entropy and Closed String Tachyons,''
  JHEP {\bf 0701}, 090 (2007)
  [hep-th/0611035].
  %%CITATION = HEP-TH/0611035;%%

%\cite{Klebanov:2007ws}
\bibitem{Klebanov:2007ws}
  I.~R.~Klebanov, D.~Kutasov and A.~Murugan,
  ``Entanglement as a probe of confinement,''
  Nucl.\ Phys.\ B {\bf 796}, 274 (2008)
  [arXiv:0709.2140 [hep-th]].
  %%CITATION = ARXIV:0709.2140;%%
%\cite{Pakman:2008ui}


\bibitem{Pakman:2008ui}
  A.~Pakman and A.~Parnachev,
  ``Topological Entanglement Entropy and Holography,''  JHEP {\bf 0807}, 097 (2008)  [arXiv:0805.1891 [hep-th]].
  %%CITATION = ARXIV:0805.1891;%


%\cite{Ogawa:2011fw}
\bibitem{Ogawa:2011fw}
  N.~Ogawa and T.~Takayanagi,
  ``Higher Derivative Corrections to Holographic Entanglement Entropy for AdS Solitons,''  JHEP {\bf 1110}, 147 (2011)  [arXiv:1107.4363 [hep-th]].  %%CITATION = ARXIV:1107.4363;%%




%\cite{Schwimmer:2008yh}
\bibitem{Schwimmer:2008yh}
  A.~Schwimmer and S.~Theisen,
  ``Entanglement Entropy, Trace Anomalies and Holography,''
  Nucl.\ Phys.\ B {\bf 801}, 1 (2008)
  [arXiv:0802.1017 [hep-th]].
  %%CITATION = ARXIV:0802.1017;%%

%\cite{Hertzberg:2010uv}
\bibitem{Hertzberg:2010uv}
  M.~P.~Hertzberg and F.~Wilczek,
  ``Some Calculable Contributions to Entanglement Entropy,''
  Phys.\ Rev.\ Lett.\  {\bf 106}, 050404 (2011)
  [arXiv:1007.0993 [hep-th]].
  %%CITATION = ARXIV:1007.0993;%%

\bibitem{Grover}
T.~Grover, A.~M.~Turner and A.~Vishwanath, ``Entanglement Entropy of Gapped Phases
and Topological Order in Three dimensions,'' Phys. Rev. B {\bf 84}, 195120 (2011)
[arXiv:1108.4038v1].




%\cite{Myers:2010xs}
\bibitem{Myers:2010xs}
  R.~C.~Myers and A.~Sinha,
  ``Seeing a c-theorem with holography,''
  Phys.\ Rev.\ D {\bf 82}, 046006 (2010)
  [arXiv:1006.1263 [hep-th]].
  %%CITATION = ARXIV:1006.1263;%%

%\cite{Myers:2010tj}
\bibitem{Myers:2010tj}
  R.~C.~Myers and A.~Sinha,
  ``Holographic c-theorems in arbitrary dimensions,''
  JHEP {\bf 1101}, 125 (2011)
  [arXiv:1011.5819 [hep-th]].
  %%CITATION = ARXIV:1011.5819;%%


%\cite{Liu:2012ee}
\bibitem{Liu:2012ee}
  H.~Liu and M.~Mezei,
  ``A refinement of entanglement entropy and the number of degrees of freedom,''
  [arXiv:1202.2070 [hep-th]].
  %%CITATION = ARXIV:1202.2070;%%

%\cite{Casini:2012ei}
\bibitem{Casini:2012ei}
  H.~Casini and M.~Huerta,
  ``On the RG running of the entanglement entropy of a circle,''
  Phys.\ Rev.\ D {\bf 85}, 125016 (2012)  [arXiv:1202.5650 [hep-th]].
  %%CITATION = ARXIV:1202.5650;%%

%\cite{Klebanov:2012yf}
\bibitem{Klebanov:2012yf}
  I.~R.~Klebanov, T.~Nishioka, S.~S.~Pufu and B.~R.~Safdi,
  ``On Shape Dependence and RG Flow of Entanglement Entropy,''
  JHEP {\bf 1207}, 001 (2012)  [arXiv:1204.4160 [hep-th]].
  %%CITATION = ARXIV:1204.4160;%%


%\cite{Myers:2012ed}
\bibitem{Myers:2012ed}
  R.~C.~Myers and A.~Singh,
  ``Comments on Holographic Entanglement Entropy and RG Flows,''
  [arXiv:1202.2068 [hep-th]].
  %%CITATION = ARXIV:1202.2068;%%


\bibitem{deconfined}
T. Senthil, A. Vishwanath, L. Balents, S. Sachdev, and M. P. A. Fisher, `` `Deconfined' quantum critical points,"  Science 303, 1490 (2004); cond-mat/0311326

 T. Senthil, L. Balents, S. Sachdev, A. Vishwanath, and M. P. A. Fisher, ``Quantum criticality beyond the Landau-Ginzburg-Wilson paradigm," Phys.\ Rev.\  B {\bf  70}, 144407 (2004); cond-mat/0312617.
 
 
%\cite{Zamolodchikov:1986gt}
\bibitem{Zamolodchikov:1986gt}
  A.~B.~Zamolodchikov,
  ``Irreversibility of the Flux of the Renormalization Group in a 2D Field Theory,''
  JETP Lett.\  {\bf 43}, 730 (1986)
  [Pisma Zh.\ Eksp.\ Teor.\ Fiz.\  {\bf 43}, 565 (1986)].
  %%CITATION = JTPLA,43,730;%%


%\cite{Komargodski:2011vj}
\bibitem{Komargodski:2011vj}
  Z.~Komargodski and A.~Schwimmer,
  ``On Renormalization Group Flows in Four Dimensions,''
  JHEP {\bf 1112}, 099 (2011)
  [arXiv:1107.3987 [hep-th]].
  %%CITATION = ARXIV:1107.3987;%%

Z.~Komargodski,
  ``The Constraints of Conformal Symmetry on RG Flows,''
  arXiv:1112.4538 [hep-th].
  %%CITATION = ARXIV:1112.4538;%%


%\cite{Jafferis:2011zi}
\bibitem{Jafferis:2011zi}
  D.~L.~Jafferis, I.~R.~Klebanov, S.~S.~Pufu and B.~R.~Safdi,
  ``Towards the F-Theorem: N=2 Field Theories on the Three-Sphere,''
  JHEP {\bf 1106}, 102 (2011)
  [arXiv:1103.1181 [hep-th]].
  %%CITATION = ARXIV:1103.1181;%%



%\cite{Fursaev:2006ih}
\bibitem{Fursaev:2006ih}
  D.~V.~Fursaev,
  ``Proof of the holographic formula for entanglement entropy,''
  JHEP {\bf 0609}, 018 (2006)
  [hep-th/0606184].
  %%CITATION = HEP-TH/0606184;%%


%\cite{Solodukhin:2008dh}
\bibitem{Solodukhin:2008dh}
  S.~N.~Solodukhin,
  ``Entanglement entropy, conformal invariance and extrinsic geometry,''  Phys.\ Lett.\ B {\bf 665}, 305 (2008)  [arXiv:0802.3117 [hep-th]].  %%CITATION = ARXIV:0802.3117;%%

%\cite{Anselmi:1997ys}
\bibitem{Anselmi:1997ys}
  D.~Anselmi, J.~Erlich, D.~Z.~Freedman and A.~A.~Johansen,
  ``Positivity constraints on anomalies in supersymmetric gauge theories,''  Phys.\ Rev.\ D {\bf 57}, 7570 (1998)  [hep-th/9711035].  %%CITATION = HEP-TH/9711035;%%


%\cite{Brigante:2008gz}
\bibitem{Brigante:2008gz}
  M.~Brigante, H.~Liu, R.~C.~Myers, S.~Shenker and S.~Yaida,
  ``The Viscosity Bound and Causality Violation,''
  Phys.\ Rev.\ Lett.\  {\bf 100}, 191601 (2008)
  [arXiv:0802.3318 [hep-th]];
  %%CITATION = ARXIV:0802.3318;%%
  ``Viscosity Bound Violation in Higher Derivative Gravity,''
  Phys.\ Rev.\ D {\bf 77}, 126006 (2008)
  [arXiv:0712.0805 [hep-th]].
  %%CITATION = ARXIV:0712.0805;%%

%\cite{Buchel:2009tt}
\bibitem{Buchel:2009tt}
  A.~Buchel and R.~C.~Myers,
  ``Causality of Holographic Hydrodynamics,''
  JHEP {\bf 0908}, 016 (2009)
  [arXiv:0906.2922 [hep-th]].
  %%CITATION = ARXIV:0906.2922;%%

%\cite{Cai:2001dz}
\bibitem{Cai:2001dz}
  R.~G.~Cai,
  ``Gauss-Bonnet black holes in AdS spaces,''
  Phys.\ Rev.\ D {\bf 65}, 084014 (2002)
  [hep-th/0109133].
  %%CITATION = HEP-TH/0109133;%%

 
%\cite{Swingle:2011mk}
\bibitem{Swingle:2011mk}
  B.~Swingle and T.~Senthil,
  ``Universal crossovers between entanglement entropy and thermal entropy,''
  [arXiv:1112.1069 [cond-mat.str-el]].
  %%CITATION = ARXIV:1112.1069;%%


  %\cite{Hartnoll:2008vx}
\bibitem{Hartnoll:2008vx}
  S.~A.~Hartnoll, C.~P.~Herzog and G.~T.~Horowitz,
  ``Building a Holographic Superconductor,''
  Phys.\ Rev.\ Lett.\  {\bf 101}, 031601 (2008)
  [arXiv:0803.3295 [hep-th]].
  %%CITATION = ARXIV:0803.3295;%%

%\cite{Horowitz:2010jq}
\bibitem{Horowitz:2010jq}
  G.~T.~Horowitz and B.~Way,
  ``Complete Phase Diagrams for a Holographic Superconductor/Insulator System,''
  JHEP {\bf 1011}, 011 (2010),
  [arXiv:1007.3714 [hep-th]].
  %%CITATION = ARXIV:1007.3714;%%

 %\cite{Hartnoll:2011fn}
\bibitem{Hartnoll:2011fn}
  S.~A.~Hartnoll,
  ``Horizons, holography and condensed matter,''
  [arXiv:1106.4324 [hep-th]].
  %%CITATION = ARXIV:1106.4324;%%

%\cite{Hubeny:2012ry}
\bibitem{Hubeny:2012ry}
  V.~E.~Hubeny,
  ``Extremal surfaces as bulk probes in AdS/CFT,''
  [arXiv:1203.1044 [hep-th]].
  %%CITATION = ARXIV:1203.1044;%%


%\cite{Cvetic:1999xp}
\bibitem{Cvetic:1999xp}
  M.~Cvetic, M.~J.~Duff, P.~Hoxha, J.~T.~Liu, H.~Lu, J.~X.~Lu, R.~Martinez-Acosta and C.~N.~Pope {\it et al.},
  ``Embedding AdS black holes in ten-dimensions and eleven-dimensions,''
  Nucl.\ Phys.\ B {\bf 558}, 96 (1999)
  [hep-th/9903214].
  %%CITATION = HEP-TH/9903214;%%



\bibitem{IC}
M.~Pawlowski, T.~Paterek, D.~Kaszlikowski, V.~Scarani,
A.~Winter, and M.~Zukowski,
``Information Causality as a Physical Principle,"
Nature, 461, 1101 (2009) [arXiv:0905.2292 [quant-ph]];

\bibitem{IC1}
 L.~-Y.~Hsu, I-C.~Yu and F.~-L.~Lin,
  ``Information Causality and Noisy Computations,''
  Phys.\ Rev.\ A {\bf 84}, 042319 (2011)
  [arXiv:1010.3419 [quant-ph]].
  %%CITATION = ARXIV:1010.3419;%%



\bibitem{GKPW}
%\cite{Gubser:1998bc}
  S.~S.~Gubser, I.~R.~Klebanov and A.~M.~Polyakov,
  ``Gauge theory correlators from non-critical string theory,''
  Phys.\ Lett.\  B {\bf 428}, 105 (1998)
  [arXiv:hep-th/9802109].
  %%CITATION = PHLTA,B428,105;%%

  E.~Witten,
  ``Anti-de Sitter space and holography,''
  Adv.\ Theor.\ Math.\ Phys.\  {\bf 2}, 253 (1998)
  [arXiv:hep-th/9802150].
  %%CITATION = 00203,2,253;%%


%\cite{Hayden:2011ag}
\bibitem{Hayden:2011ag}
  P.~Hayden, M.~Headrick and A.~Maloney, ``Holographic Mutual Information is Monogamous,''
 [arXiv:1107.2940 [hep-th]].
  %%CITATION = ARXIV:1107.2940;%%

\bibitem{SRE1}
X.~Chen, Z.-C. Gu, X.-G. Wen,
``Classification of Gapped Symmetric Phases in 1D Spin Systems,"
Phys.\ Rev.\ B {\bf 83}, 035107 (2011) [arXiv:1008.3745[cond-mat]]

\bibitem{SRE2}
X.~Chen, Z.-C. Gu, Z.-X. Liu, X.-G. Wen,
``Symmetry protected topological orders and the group cohomology of their symmetry group,"
[arXiv:1106.4772[cond-mat]].

\bibitem{SRE3}
X.~Chen, Z.-C. Gu, X.-G. Wen,
``Towards a complete classification of 1D gapped quantum phases in interacting spin systems,"
[arXiv:1103.3323[cond-mat]].

\bibitem{SRE4}
F.~Pollmann, E.~ Berg, A.~ M.~Turner, M.~ Oshikawa,
``Entanglement spectrum of a topological phase in one dimension", Phys.\ Rev.\ B.\ {\bf 81}, 064439 (2010) [arXiv:0910.1811[cond-mat]];
``Symmetry protection of topological order in one-dimensional quantum spin systems", [arXiv:0909.4059[cond-mat]].

\bibitem{SRE5}
N.~Schuch, D.~Perez-Garcia, I.~Cirac,
``Classifying quantum phases using Matrix Product States and PEPS," Phys.\ Rev.\ B.\ {\bf  84}, 165139 (2011) [ arXiv:1010.3732[cond-mat]].




%\cite{Swingle:2009bg}
\bibitem{Swingle:2009bg}
  B.~Swingle, ``Entanglement Renormalization and Holography,''
 [arXiv:0905.1317 [cond-mat.str-el]].
  %%CITATION = ARXIV:0905.1317;%%

 \bibitem{EvenblyVidal}
 G.~Evenbly and G.~Vidal, ``Tensor network states and geometry", [arXiv:1106.1082[quant-ph]].

 \bibitem{MERA}
 G.~Vidal, ``Entanglement renormalization", Phys.\ Rev.\ Lett.\ 98, 070201 (2007) [arXiv:cond-mat/0512165];  ``A class of quantum many-body states that can be efficiently simulated", Phys. Rev. Lett. 101, 110501 (2008), [ arXiv:quant-ph/0610099].


 \bibitem{Evenbly}
 G.~Evenbly, ``Foundations and Applications of Entanglement Renormalization," [ arXiv:1109.5424 [quant-ph]].

\bibitem{qsrg1}
F.~Verstraete, J.~I.~Cirac, J.~I.~Latorre, E.~Rico, M.~M.~Wolf, ``Renormalization group transformations on quantum states,"  Phys.\ Rev.\ Lett.\ 94 (2005) 140601 [quant-ph/0410227v1].

\bibitem{qsrg2}
X.~Chen, Z.-C.~Gu, X.-G.~Wen, ``Local unitary transformation, long-range quantum entanglement, wave function renormalization, and topological order,"   Phys.\ Rev.\ B 82, 155138 (20 10) [arXiv:1004.3835 [cond-mat.str-el]].


\bibitem{qsrg3}
C.-Y.~Huang, F.-L.~Lin, ``Topological order and degenerate singular value spectrum in two-dimensional dimerized quantum Heisenberg model", Phys.\ Rev.\  B.\ {bf 84}, 125110 (2011) [arXiv:1104.1139 [cond-mat.str-el]].

\bibitem{qsrg4}
C.-Y.~Huang, X.~Chen, F.-L.~Lin, ``Symmetry Protected Quantum State Renormalization",  [arXiv:1303.4190 [cond-mat.str-el]].




\end{thebibliography}
\end{document}